\documentclass[amssymb,amsmath,aps,pre,showpacs]{revtex4}
\usepackage{graphicx}

\begin{document}

\title{Ginzburg-Landau theory of the liquid-solid interface and nucleation
for hard-spheres}
\author{James F. Lutsko}
\affiliation{Center for Nonlinear Phenomena and Complex Systems, Universit\'{e} Libre de
Bruxelles, C.P. 231, Blvd. du Triomphe, 1050 Brussels, Belgium}
\pacs{64.10.+h,68.08.-p}
\date{\today }

\begin{abstract}
The Ginzburg-Landau free energy functional for hard-spheres is constructed
using the Fundamental Measure Theory approach to Density Functional Theory
as a starting point. The functional is used to study the liquid-fcc solid
planer interface and the properties of small solid clusters nucleating
within a liquid. The surface tension for planer interfaces agrees well with
simulation and it is found that the properties of the solid clusters are
consistent with classical nucleation theory.
\end{abstract}

\maketitle

\section{Introduction}

The nature of the liquid-solid interface is of both practical and
theoretical significance. From a practical point of view, the interfacial
surface tension is, along with the bulk free energies, one of the
determining factors controlling nucleation of crystals from solution and
plays as well an important role in the closely related phenomena of wetting.
However, a complete description of the liquid-solid interface based on
fundamental considerations remains a challenging theoretical problem since
it requires knowledge of the spatial -dependence of the free energy in
highly inhomogeneous systems. Much can be learned from more phenomenological
approaches such as phase-field theory and model free energy functionals but
a detailed relation of interfacial properties to molecular interaction
models requires more fundamental approaches. On the other hand, very
detailed approaches such as molecular dynamics and Monte Carlo computer
simulation can provide direct microscopic information about interfaces, but
theory is required in order to understand the results obtained.

Classical density functional theory (DFT) is based on the fact that the
Helmholtz free energy is a unique functional of the average density profile%
\cite{EvansDFT},\cite{HansenMcdonald}. It has been used to study
liquid-solid interfaces - both coexistence and wetting - almost since its
inception\cite{EvansDFT},\cite{OxtobyHaymet1},\cite{OxtobyHaymet2}. The most
commonly studied system is that of hard-spheres since many density
functional theories work best in this case. Although somewhat artificial,
the hard-sphere interaction plays an important role in equilibrium
statistical mechanics since more realistic pair interactions can be
described based on perturbation theory about the hard-sphere interaction\cite%
{HansenMcdonald} or by developments within DFT inspired by perturbation
theory\cite{CurtinAschroftPert}. However, the theories which most accurately
described the bulk liquid and solid phases for hard-spheres suffer from a
serious technical deficiency in that they allow for configurations in which
the spheres overlap - a situation which is excluded on physical grounds and
which should manifest itself as a divergence in the free energy for
configurations in which the spheres touch\cite{OLW}. Thus, the liquid-solid
interface can only be studied with these theories if either the allowed
densities are artificially restricted, as in \cite{Curtin-Interface, Gast1},
or if ad hoc modifications are made to the DFT so as to create the required
divergence at overlap\cite{OLW,OLW2}. In recent years, a new class of DFT
models, generally known as Fundamental Measure Theory (FMT), have proven
successful in a number of applications particularly involving inhomogeneous
fluids (e.g., fluids near walls)\cite{Rosenfeld1},\cite{Rosenfeld2},\cite%
{Rosenfeld_1997_1}. The FMT approach has the advantage that, by the nature
of the models used, the problem of overlapping hard-spheres is automatically
solved - overlapping spheres lead to an infinite free energy penalty as one
would expect. The goal of this paper is therefore to revisit the problem of
the liquid-solid interface for hard-spheres both to further test the
generality of the FMT\ approach and also as a preliminary step towards the
study of interfaces for more realistic systems.

Within the DFT-FMT\ framework, there is still considerable latitude in the
level of description of the physical system. A bulk liquid is characterized
by a constant average density $\rho \left( \overrightarrow{r}\right) =%
\overline{\rho }$ while a bulk solid is characterized by a spatially-varying
density which must have the symmetry of the underlying crystal lattice.
Inhomogeneous systems with either interfaces between different phases or
confined geometries necessarily have more complex density profiles involving
non-periodic spatial variations. For liquids, this does not not pose too
great a challenge but for solids, the superposition of the spatial
variations within a unit cell and the larger-scale variations coming from
the interface can lead to numerically-intensive calculations\, see e.g ref. %
\cite{OLW2}. Indeed, a recent application of the DFT-FMT approach to the
problem of hard-sphere liquid-solid coexistence has been made and serves to
illustrate the difficulty of this approach\cite{Song_FMT}. In this work, the
goal is to use a reduced description whereby a small set of order parameters
rather than the fully detailed local density\cite{Lutsko_GL}. In this sense,
the present work follows the spirit of phase field methods. However, the
free energy functional is systematically derived from the microscopic
description thus eliminating the need for phenomenological assumptions and
thus allowing for quantitative predictions.

In the next Section, the elements of DFT and its relation to the
Ginzburg-Landau theory are reviewed. The FMT\ density functionals are also
described and the use of these functionals to calculate the elements of the
GL free energy functional is presented. In Section III, the properties of
the planer liquid-solid interface are determined both by means of
parametrized profiles of the density and crystallinity and by numerical
solution of the Euler-Lagrange equations. The calculated surface tension is
shown to be in reasonable agreement with the results of molecular dynamics
and Monte Carlo simulations. The structure and free energy of small solid
clusters is also discussed. It is found that the properties of the critical
nucleus - its size and the widths of the interfacial region - are well
predicted by the results obtained for the planer interface using classical
nucleation theory. The last Section summarizes the results and discusses
possible refinements of the calculations.

\section{Theory}

\subsection{Density functional theory}

Density functional theory is based on the fact that the grand potential, the
thermodynamic free energy appropriate for a system with constant chemical
potential $\mu $, constant temperature $T$ and fixed applied one-body field $%
\phi \left( \overrightarrow{r}\right) $, can be written as%
\begin{equation}
\Omega =F\left[ \rho \right] -\int \mu \rho \left( \overrightarrow{r}\right)
d\overrightarrow{r}+\int \phi \left( \overrightarrow{r}\right) \rho \left( 
\overrightarrow{r}\right) d\overrightarrow{r}
\end{equation}%
where the first term on the right is a unique functional of the local
density. Different applied fields will give rise to different density
profiles and the fundamental theorem of DFT says that there is a one to one
correspondence between applied fields and density profiles\cite{EvansDFT},%
\cite{HansenMcdonald}. It follows from this definition that the Helmholtz
free energy, $A,$ is 
\begin{equation}
A=F\left[ \rho \right] +\int \phi \left( \overrightarrow{r}\right) \rho
\left( \overrightarrow{r}\right) d\overrightarrow{r}
\end{equation}%
so that $F\left[ \rho \right] $ is the intrinsic contribution to the
Helmholtz free energy due to the density profile. For situations in which
the applied field is not important - for example, when it only represents
the walls of a container, the second term is unimportant in the
thermodynamic limit and $F\left[ \rho \right] $ is often referred to simply
as the Helmholtz free energy.

At fixed field, temperature and chemical potential, the density must
minimize the grand potential giving the Euler-Lagrange equation%
\begin{equation}
\frac{\delta }{\delta \rho \left( \overrightarrow{r}\right) }F\left[ \rho %
\right] -\mu +\phi \left( \overrightarrow{r}\right) =0.
\end{equation}%
Given a model for the intrinsic free energy functional, $F\left[ \rho \right]
$, this gives a closed description of the system which generally takes the
form of an integral equation. Most applications of DFT make use of
parametrized density profiles so as to reduce the effort needed to determine
the density. In general, if the density is given by $\rho \left( 
\overrightarrow{r}\right) =\rho \left( \overrightarrow{r};\Gamma \right) $,
where $\rho \left( \overrightarrow{r};\Gamma \right) $ is some fixed
function of the spatial coordinates and the permeates $\Gamma =\left\{
\Gamma _{a}\right\} _{a=1}^{n_{\Gamma }}$ , then the Euler-Lagrange
equations become%
\begin{equation}
\frac{\partial }{\partial \Gamma _{a}}F\left[ \rho \right] -\mu \frac{%
\partial \overline{\rho }}{\partial \Gamma _{a}}+\int \phi \left( 
\overrightarrow{r}\right) \frac{\delta \rho \left( \overrightarrow{r};\Gamma
\right) }{\delta \Gamma _{a}}d\overrightarrow{r}=0.
\end{equation}%
where the notation indicates that $F\left[ \rho \right] =F\left( \Gamma
\right) $ is an ordinary function of the parameters. The derivatives are
understood to be evaluated holding all parameters constant except $\Gamma
_{a}$. Note that if one of the parameters corresponds to the average
density, say $\Gamma _{0}=\overline{\rho }$, and if the field is zero (or
confined to the boundaries so that it can be neglected in the thermodynamic
limit) this gives%
\begin{eqnarray}
\frac{\partial }{\partial \overline{\rho }}F\left[ \rho \right] &=&\mu
\label{thermo} \\
\frac{\partial }{\partial \Gamma _{a}}F\left[ \rho \right] &=&0,\;\;a>0. 
\nonumber
\end{eqnarray}%
The first equation is just the usual relation between the Helmholtz free
energy and the chemical potential while the second says that the Helmholtz
free energy must be stationary with respect to all of the other parameters.

For a bulk solid corresponding to a Bravais lattice with a single atom per
unit cell, the density must have the symmetry of the lattice and so takes
the form%
\begin{equation}
\rho \left( \overrightarrow{r}\right) =\sum_{n}f\left( \overrightarrow{r}-%
\overrightarrow{R}_{n}\right)  \label{d1}
\end{equation}%
where $\left\{ \overrightarrow{R}_{n}\right\} $ are the lattice vectors. In
this case, the average density is 
\begin{eqnarray}
\overline{\rho } &=&\frac{1}{V}\int \rho \left( \overrightarrow{r}\right) d%
\overrightarrow{r}  \label{d2} \\
&=&\overline{\rho }_{latt}\int_{WS}f\left( \overrightarrow{r}\right) d%
\overrightarrow{r}  \nonumber
\end{eqnarray}%
where the second integral is restricted to the Wigner-Seitz cell and where $%
\overline{\rho }_{latt}$ is the lattice density, defined as the number of
lattice points per unit volume. The integral therefore defines the
occupancy, $x_{0}= \overline{\rho }/\overline{\rho }_{latt}$: an occupancy
of one means that every lattice site is occupied, a value less than one
means that there are some vacancies, a value greater than one means that
there are some interstitials. Note that eq.(\ref{d1}) can equivalently be
written in terms of Fourier components as%
\begin{equation}
\rho \left( \overrightarrow{r}\right) =\sum_{n}\exp \left( i\overrightarrow{K%
}_{n}\cdot \overrightarrow{r}\right) \widetilde{f}\left( \overrightarrow{K}%
_{n}\right)  \label{d3}
\end{equation}%
where $\widetilde{f}\left( \overrightarrow{k}\right) $ is the Fourier
transform of $f\left( \overrightarrow{r}\right) $ and where $\left\{ 
\overrightarrow{K}_{n}\right\} $ is the set of reciprocal lattice vectors.

A typical parametrization of the density, widely used in practical
calculations, is to take $f\left( \overrightarrow{r}\right) $ to be a
Gaussian so that%
\begin{equation}
f\left( \overrightarrow{r}\right) =x_{0}\left( \frac{\alpha }{\pi }\right)
^{3/2}\exp \left( -\alpha r^{2}\right) .  \label{d4}
\end{equation}%
There are then three parameters:\ the average density, $\overline{\rho }$,
the width of the Gaussian, $\alpha $, and the lattice density $\overline{%
\rho }_{latt}$ and the occupancy is just $x_{0}=\overline{\rho }/\rho _{latt}
$. In this case, one has $\widetilde{f}\left( \overrightarrow{K}_{n}\right) =%
\overline{\rho }\exp \left( -K_{n}^{2}/4\alpha \right) $ which, together
with the fact that $\overrightarrow{K}_{0}=0$ shows that $\lim_{\alpha
\rightarrow 0}\rho \left( \overrightarrow{r};\Gamma \right) =\overline{\rho }
$ so that the parametrization of the density encompasses both the Gaussian
approximation for the solid and the uniform liquid. For this reason, it is
common to take the value of the first non-trivial Fourier component to be a
measure of the ''crystallinity'', denoted $m$, giving%
\begin{equation}
m=\exp \left( -K_{1}^{2}/4\alpha \right)
\end{equation}%
so that $m=0$ corresponds to a uniform fluid and $m=1$ to an infinitely
localized solid. Real solids have values of $m$ which are close to, but
always less than, one.

To study interfacial properties, it is necessary to allow for spatial
variation in both the average density and in the crystallinity. Here, we
follow Ohensorge, et al\cite{OLW} and allow $\overline{\rho }$ and $\alpha $
to vary giving%
\begin{equation}
\rho \left( \overrightarrow{r}\right) =\left( \overline{\rho }\left( 
\overrightarrow{r}\right) /\rho _{latt}\right) \left( \frac{\alpha \left( 
\overrightarrow{r}\right) }{\pi }\right) ^{3/2}\sum_{n}\exp \left( -\alpha
\left( \overrightarrow{r}\right) \left( \overrightarrow{r}-\overrightarrow{R}%
_{n}\right) ^{2}\right)  \label{r1}
\end{equation}%
or%
\begin{equation}
\rho \left( \overrightarrow{r}\right) =\overline{\rho }\left( 
\overrightarrow{r}\right) \sum_{n}\exp \left( i\overrightarrow{K}_{n}\cdot 
\overrightarrow{r}\right) \exp \left( -K_{n}^{2}/4\alpha \left( 
\overrightarrow{r}\right) \right)
\end{equation}%
depending on which form of the bulk density is used as a basis for the
generalization. These expressions are obviously not equivalent although it
will turn out below that within the Ginzburg-Landau framework, the
differences are unimportant. An alternative parametrization used by Haymet
and Oxtoby\cite{OxtobyHaymet1},\cite{OxtobyHaymet2} is%
\begin{equation}
\rho \left( \overrightarrow{r}\right) =\delta \rho \left( \overrightarrow{r}%
\right) +\left( \frac{\alpha \left( \overrightarrow{r}\right) }{\pi }\right)
^{3/2}\sum_{n}\exp \left( -\alpha \left( \overrightarrow{r}\right) \left( 
\overrightarrow{r}-\overrightarrow{R}_{n}\right) ^{2}\right)
\end{equation}%
or%
\begin{equation}
\rho \left( \overrightarrow{r}\right) =\delta \rho \left( \overrightarrow{r}%
\right) +\overline{\rho }_{latt}\sum_{n}\exp \left( i\overrightarrow{K}%
_{n}\cdot \overrightarrow{r}\right) \exp \left( -K_{n}^{2}/4\alpha \left( 
\overrightarrow{r}\right) \right)
\end{equation}%
where it is assumed that $\overline{\rho }_{s}=\overline{\rho }_{latt}$ in
the solid. One difficulty with this form is that since $\delta \rho \left( 
\overrightarrow{r}\right) $ must be allowed to be negative (the liquid is
less dense than the solid)\ there is the unphysical possibility that $\rho
\left( \overrightarrow{r}\right) <0$ for some points $\overrightarrow{r}$.
The form given in eq.(\ref{r1}) is positive definite thus avoiding this
problem.

\subsection{Fundamental Measure Theory}

The original form of Fundamental Measure Theory as given by Rosenfeld\cite%
{Rosenfeld1},\cite{Rosenfeld2} is in some sense a development of scaled
particle theory. This was further extended by Tarazona, Rosenfeld and others
using the important requirement that the known, exact free energy functional
be recovered in the one-dimensional limit of the theory\cite{Tarazona_1997_2}%
,\cite{Rosenfeld_1997_1},\cite{tarazona_2000_1}. The resulting class of
theories has proven successful at describing inhomogeneous hard-sphere
fluids including the hard-sphere solid. The theory involves a number of
local variables, $n_{\alpha }\left( \overrightarrow{r}_{1}\right) $, which
are linear functionals of the local density of the form 
\begin{equation}
n_{\alpha }\left( \overrightarrow{r}_{1}\right) =\int d\overrightarrow{r}%
_{2}\;\rho \left( \overrightarrow{r}_{2}\right) w_{\alpha }\left( 
\overrightarrow{r}_{12}\right) .
\end{equation}%
The set of weights $w_{\alpha }\left( \overrightarrow{r}_{12}\right) $
include a simple step function $\Theta \left( \frac{\sigma }{2}-r\right) $
which serves to define a local packing fraction%
\begin{equation}
\eta \left( \overrightarrow{r}_{1}\right) =\int d\overrightarrow{r}%
_{2}\;\rho \left( \overrightarrow{r}_{2}\right) \Theta \left( \frac{\sigma }{%
2}-r_{12}\right)
\end{equation}%
as evidenced by the fact that in the uniform limit, $\rho \left( 
\overrightarrow{r}_{2}\right) \rightarrow \overline{\rho }$, one has $\eta
\left( \overrightarrow{r}_{1}\right) =\frac{\pi }{6}\overline{\rho }\sigma
^{3}$ which is the usual definition of the hard-sphere packing fraction. All
of the remaining weighting functions are tensorial dyadics of the form $%
\widehat{r}\widehat{r}...\widehat{r}\delta \left( r-\frac{\sigma }{2}\right) 
$ and the resulting variables are written generically as%
\begin{equation}
T_{ij...l}\left( \overrightarrow{r}_{1}\right) =\int d\overrightarrow{r}%
_{2}\;\widehat{r}_{12,i}\widehat{r}_{12,j}...\widehat{r}_{12,l}\delta \left(
r_{12}-\frac{\sigma }{2}\right) \rho \left( \overrightarrow{r}_{2}\right) .
\end{equation}%
It will be useful to give simpler names for the first two of these
quantities, namely%
\begin{eqnarray}
s\left( \overrightarrow{r}_{1}\right) &=&\int d\overrightarrow{r}%
_{2}\;\delta \left( r_{12}-\frac{\sigma }{2}\right) \rho \left( 
\overrightarrow{r}_{2}\right) \\
v_{i}\left( \overrightarrow{r}_{1}\right) &=&\int d\overrightarrow{r}_{2}\;%
\widehat{r}_{12,i}\delta \left( r_{12}-\frac{\sigma }{2}\right) \rho \left( 
\overrightarrow{r}_{2}\right)  \nonumber
\end{eqnarray}%
where the names stand for ''scalar'' and ''vector'' respectively.

The Helmholtz free energy functional is written as%
\begin{equation}
F\left[ \rho \right] =F_{id}\left[ \rho \right] +F_{ex}\left[ \rho \right]
\end{equation}%
where the ideal part of the free energy is 
\begin{equation}
\beta F_{id}\left[ \rho \right] =\int d\overrightarrow{r}\;\left( \rho
\left( \overrightarrow{r}\right) \ln \rho \left( \overrightarrow{r}\right)
-\rho \left( \overrightarrow{r}\right) \right)
\end{equation}%
with $\beta = 1/(k_{B}T)$, and the excess contribution is written in the
FMT\ as the integral of a function of the local variables 
\begin{equation}
\beta F_{ex}\left[ \rho \right] =\int d\overrightarrow{r}\;\beta \phi \left(
\left\{ n_{\alpha }\left( \overrightarrow{r}\right) \right\} \right)
\end{equation}%
which is usually expressed as%
\begin{equation}
\phi =\phi _{1}+\phi _{2}+\phi _{3}
\end{equation}%
with%
\begin{eqnarray}
\beta \phi _{1} &=&-\frac{1}{\pi \sigma ^{2}}s\left( \overrightarrow{r}%
\right) \ln \left( 1-\eta \left( \overrightarrow{r}\right) \right) \\
\beta \phi _{2} &=&\frac{1}{2\pi \sigma }\frac{s^{2}\left( \overrightarrow{r}%
\right) -v^{2}\left( \overrightarrow{r}\right) }{\left( 1-\eta \left( 
\overrightarrow{r}\right) \right) }  \nonumber
\end{eqnarray}%
The form of $\phi_{3}$ depends on the particular version of FMT. Here, three
common theories will be considered. The first theory was proposed by
Rosenfeld et al\cite{Rosenfeld_1997_1} and is perhaps the simplest form of
FMT\ capable of giving a good description of the hard-sphere solid%
\begin{equation}
\beta \phi _{3}^{RLST}=\frac{\frac{1}{3}s^{3}\left( \overrightarrow{r}%
\right) }{8\pi \left( 1-\eta \left( \overrightarrow{r}\right) \right) ^{2}}%
\left( 1-\frac{v^{2}\left( \overrightarrow{r}\right) }{s^{2}\left( 
\overrightarrow{r}\right) }\right) ^{3}.
\end{equation}%
The second is the theory of Tarazona\cite{tarazona_2000_1} which makes use
of a tensor variable%
\begin{equation}
\beta \phi _{3}^{T}=\frac{3}{16\pi }\frac{1}{\left( 1-\eta \left( 
\overrightarrow{r}\right) \right) ^{2}}\left( \overrightarrow{v}\left( 
\overrightarrow{r}\right) \cdot \overleftrightarrow{T}\left( \overrightarrow{%
r}\right) \cdot \overrightarrow{v}\left( \overrightarrow{r}\right) -s\left( 
\overrightarrow{r}\right) v^{2}\left( \overrightarrow{r}\right) -Tr\left( 
\overleftrightarrow{T}^{3}\left( \overrightarrow{r}\right) \right) +s\left( 
\overrightarrow{r}\right) Tr\left( \overleftrightarrow{T}^{2}\left( 
\overrightarrow{r}\right) \right) \right) .
\end{equation}%
Both of these theories have the property that they reduce to the
Percus-Yevik approximation for the liquid. The third theory builds in the
more accurate Carnahan-Starling equation of state via a heuristic
modification of the Tarazona theory\cite{tarazona_2002_1},\cite{WhiteBear}.
It is commonly known as the ''White Bear''\ functional and is given by 
\begin{equation}
\beta \phi _{3}^{WB}=\frac{2}{3}\frac{\eta \left( \overrightarrow{r}\right)
+\left( 1-\eta \left( \overrightarrow{r}\right) \right) ^{2}\ln \left(
1-\eta \left( \overrightarrow{r}\right) \right) }{\eta ^{2}\left( 
\overrightarrow{r}\right) }\phi _{3}^{T}.
\end{equation}
All three of these theories have deficiencies. The RLST theory incorporates
the Percus-Yevik approximation for the liquid which is not very accurate at
the density of liquid-solid coexistence. To the extent that the RLST theory
gives a good description of liquid-solid coexistence (see below), it is
because it gets the liquid and solid ``equally wrong''. The Tarazona theory
also reduces to the Percus-Yevik approximation for the homogeneous fluid but
it gives a better description of the properties of the homogeneous solid
leading to a poor description of coexistence\cite{tarazona_2002_1}. For this
reason, this theory has not been used in the present investigation. The
White Bear (WB) functional gives an improved description of the dense fluid
by incorporating the Carnahan-Starling equation of state in an ad hoc
fashion. As a result, the implied pair distribution function for the fluid,
obtained via the Ornstein-Zernike equation, will not vanish in the core
region as it should. Nevertheless, the free energy does diverge for
overlapping hard-spheres as in the other forms of FMT. The conclusion is
that the RLST is probably the best theory in terms of the formal properties
of the free energy while the WB may be expected to be the better in terms of
quantitative results.

\subsection{Ginzburg-Landau theory}

If it can be assumed that the order parameters vary slowly over atomic
length scales, a simplified free energy functional can be rigorously derived
from the exact free energy functional by means of a gradient expansion in
the order parameters. When carried out to second order, the result takes the
form of a phenomenological Ginzburg-Landau free energy functional\cite%
{EvansDFT},\cite{OxtobyHaymet1},\cite{OxtobyHaymet2},\cite{Lowen1},\cite%
{Lowen2},\cite{Lutsko_GL},%
\begin{equation}
\beta \Omega _{GL}\left[ \Gamma \right] =\int d\overrightarrow{R}\;\left[ 
\frac{1}{V}\beta F\left( \Gamma \left( \overrightarrow{R}\right) \right)
-\beta \mu \overline{\rho }\left( \Gamma \left( \overrightarrow{R}\right)
\right) +\frac{1}{2}K_{ij}^{ab}\left( \Gamma \left( \overrightarrow{R}%
\right) \right) \frac{\partial \Gamma _{a}\left( \overrightarrow{R}\right) }{%
\partial R_{i}}\frac{\partial \Gamma _{b}\left( \overrightarrow{R}\right) }{%
\partial R_{j}}\right]
\end{equation}%
where the mean field term $\beta F\left( \Gamma \right) $ is the free energy
of a bulk system evaluated with order parameters $\Gamma $ and%
\[
\overline{\rho }\left( \Gamma \right) =\int \rho \left( \overrightarrow{r}%
_{1};\Gamma \right) d\overrightarrow{r}_{1}. 
\]%
The coefficient of the gradient term is 
\begin{equation}
K_{ij}^{ab}\left( \Gamma \right) =\frac{1}{2V}\int d\overrightarrow{r}_{1}d%
\overrightarrow{r}_{2}\;r_{12i}r_{12j}c_{2}\left( \overrightarrow{r}_{1},%
\overrightarrow{r}_{2};\Gamma \right) \frac{\partial \rho \left( 
\overrightarrow{r}_{1};\Gamma \right) }{\partial \Gamma _{a}}\frac{\partial
\rho \left( \overrightarrow{r}_{2};\Gamma \right) }{\partial \Gamma _{b}}.
\label{K}
\end{equation}%
where the direct correlation function is also evaluated for a bulk system
with constant order parameters and is determined from the free energy via
the standard relation%
\begin{equation}
c_{2}\left( \overrightarrow{r}_{1},\overrightarrow{r}_{2};\Gamma \right) =-%
\frac{\delta ^{2}\beta F_{ex}\left[ \rho \right] }{\delta \rho \left( 
\overrightarrow{r}_{1};\Gamma \right) \delta \rho \left( \overrightarrow{r}%
_{2};\Gamma \right) }.
\end{equation}%
As discussed in ref.\cite{Lutsko_GL}, this expression is derived under the
assumption that the lattice structure is held fixed. This precludes the use
of any characteristics of the lattice, such as the lattice constants or the
primitive lattice vectors, as order parameters. Nevertheless, it is possible
to explore the liquid-solid interface since, as discussed above, this can be
done for a fixed lattice structure.

When the underlying lattice has cubic symmetry, as will be the case here,
and in a coordinate system aligned with the principle axes of the lattice
structure, the symmetry under 90 degree rotations around the axes implies
that the coefficient of the gradient term can be written as%
\begin{equation}
K_{ij}^{ab}=g_{ab}\delta _{ij}+h_{ab}\left( \delta _{ix}\delta _{jy}+\delta
_{iy}\delta _{jx}+\delta _{ix}\delta _{jz}+\delta _{iz}\delta _{jx}+\delta
_{iy}\delta _{jz}+\delta _{iz}\delta _{jy}\right)
\end{equation}%
where 
\begin{eqnarray}
g_{ab}\left( \Gamma \right) &=&\frac{1}{6V}\int d\overrightarrow{r}_{1}d%
\overrightarrow{r}_{2}\;r_{12}^{2}c_{2}\left( \overrightarrow{r}_{1},%
\overrightarrow{r}_{2};\Gamma \right) \frac{\partial \rho \left( 
\overrightarrow{r}_{1};\Gamma \right) }{\partial \Gamma _{a}}\frac{\partial
\rho \left( \overrightarrow{r}_{2};\Gamma \right) }{\partial \Gamma _{b}}
\label{gh} \\
h_{ab}\left( \Gamma \right) &=&\frac{1}{4V}\int d\overrightarrow{r}_{1}d%
\overrightarrow{r}_{2}\;r_{12x}r_{12y}c_{2}\left( \overrightarrow{r}_{1},%
\overrightarrow{r}_{2};\Gamma \right) \frac{\partial \rho \left( 
\overrightarrow{r}_{1};\Gamma \right) }{\partial \Gamma _{a}}\frac{\partial
\rho \left( \overrightarrow{r}_{2};\Gamma \right) }{\partial \Gamma _{b}}. 
\nonumber
\end{eqnarray}

To apply this formalism to the FMT, note that the direct correlation
function takes the form%
\begin{equation}
c_{2}\left( \overrightarrow{r}_{1},\overrightarrow{r}_{2};\Gamma \right)
=-\sum_{\alpha ,\beta }\int d\overrightarrow{r}\;\frac{\partial ^{2}\phi
\left( \left\{ n_{\alpha }\left( \overrightarrow{r}\right) \right\} \right) 
}{\partial n_{\alpha }\partial n_{\beta }}w_{\alpha }\left( \overrightarrow{r%
}-\overrightarrow{r}_{1}\right) w_{\beta }\left( \overrightarrow{r}-%
\overrightarrow{r}_{2}\right)  \label{c2}
\end{equation}%
At first, it appears that the evaluation of the matrix $K_{ij}^{ab}$ will
involve triple volume integrals making it extremely expensive to calculate.
However, it is in fact possible to arrange the calculate so that it takes no
more effort to calculate $K_{ij}^{ab}$ than is required to calculate the
free energy. To do this, note that the calculation requires evaluation of
the integral%
\begin{equation}
\int d\overrightarrow{r}_{1}d\overrightarrow{r}_{2}\;r_{12i}r_{12j}w_{\alpha
}\left( \overrightarrow{r}-\overrightarrow{r}_{1}\right) w_{\beta }\left( 
\overrightarrow{r}-\overrightarrow{r}_{2}\right) \frac{\partial \rho \left( 
\overrightarrow{r}_{1};\Gamma \right) }{\partial \Gamma _{a}}\frac{\partial
\rho \left( \overrightarrow{r}_{2};\Gamma \right) }{\partial \Gamma _{b}}
\end{equation}%
which can be simplified by introducing some additional functionals of the
density. Specifically, let%
\begin{eqnarray}
n_{\alpha }^{i}\left( \overrightarrow{r}\right) &=&\int d\overrightarrow{r}%
_{1}\;r_{1i}w_{\alpha }\left( \overrightarrow{r}-\overrightarrow{r}%
_{1}\right) \rho \left( \overrightarrow{r}_{1};\Gamma \right) \\
n_{\alpha }^{ij}\left( \overrightarrow{r}\right) &=&\int d\overrightarrow{r}%
_{1}\;r_{1i}r_{1j}w_{\alpha }\left( \overrightarrow{r}-\overrightarrow{r}%
_{1}\right) \rho \left( \overrightarrow{r}_{1};\Gamma \right)  \nonumber
\end{eqnarray}%
so that the full expression for $K_{ij}^{ab}$ becomes%
\begin{equation}
K_{ij}^{ab}\left( \Gamma \right) =-\frac{1}{2V}\sum_{\alpha ,\beta }\int d%
\overrightarrow{r}\;\frac{\partial ^{2}\phi \left( \left\{ n_{\alpha }\left( 
\overrightarrow{r}\right) \right\} \right) }{\partial n_{\alpha }\partial
n_{\beta }}\left( 
\begin{array}{c}
\frac{\partial n_{\alpha }^{ij}\left( \overrightarrow{r}\right) }{\partial
\Gamma _{a}}\frac{\partial n_{\beta }\left( \overrightarrow{r}\right) }{%
\partial \Gamma _{b}}+\frac{\partial n_{\alpha }\left( \overrightarrow{r}%
\right) }{\partial \Gamma _{a}}\frac{\partial n_{\beta }^{ij}\left( 
\overrightarrow{r}\right) }{\partial \Gamma _{b}} \\ 
-\frac{\partial n_{\alpha }^{i}\left( \overrightarrow{r}\right) }{\partial
\Gamma _{a}}\frac{\partial n_{\beta }^{j}\left( \overrightarrow{r}\right) }{%
\partial \Gamma _{b}}-\frac{\partial n_{\alpha }^{j}\left( \overrightarrow{r}%
\right) }{\partial \Gamma _{a}}\frac{\partial n_{\beta }^{i}\left( 
\overrightarrow{r}\right) }{\partial \Gamma _{b}}%
\end{array}%
\right) .
\end{equation}%
This is of the same structural form as the expression for the excess free
energy,  namely a spatial integral involving functions of the local density
functionals, and can therefore be evaluated as easily. Furthermore, note
that for the tensorial densities, one has%
\begin{eqnarray}
T_{ij...l}^{m}\left( \overrightarrow{r}\right) &=&\left( \frac{\sigma }{2}%
\right) T_{ij...lm}\left( \overrightarrow{r}\right) \\
T_{ij...l}^{mn}\left( \overrightarrow{r}\right) &=&\left( \frac{\sigma }{2}%
\right) ^{2}T_{ij...lmn}\left( \overrightarrow{r}\right)  \nonumber
\end{eqnarray}%
The only other quantities needed to evaluate the GL free energy functional
are $\eta ^{i}\left( \overrightarrow{r}\right) $ and $\eta ^{ij}\left( 
\overrightarrow{r}\right) $. Thus, for the RLST theory, the tensorial
quantities must be evaluated up to third order while the White Bear theory
requires the fourth order tensor as well. Explicit expressions for all of
these quantities are given in Appendix \ref{app-details}. An interesting
analytic result, proven in Appendix \ref{App-h} is that in the RLST theory $%
h_{ab}\left( \Gamma \right) =0$. Using the WB theory, it is possible that $%
h_{ab}\left( \Gamma \right) \neq 0$ but the present calculations for the
case of an FCC lattice are that the calculated values are so small that
neglect of $h_{ab}\left( \Gamma \right) $ makes no difference to the results
reported below. Thus, for all practical purposes, the GL free energy
functional is not affected by the anisotropy of the underlying lattice for
hard spheres.

\bigskip

\subsection{Numerical methods}

Many quantities, such as the density, can be evaluated either in real space
or in Fourier space. Generally, the former is more efficient in the (large $%
\alpha$) crystalline state and the latter in the (small $\alpha$) liquid
state. For the present study, both methods have been used in the calculation
of the density and of all of the reqiured density functionals (see Appendix %
\ref{app-details} for details). The implementation was checked by comparing
both methods for values of $\alpha a^{2}\sim 20$, where $a=\left( 4 / 
\overline{\rho}_{latt}\right)^{1/3}$ is the FCC lattice paramter. In all
subsequent calculations, the Fourier-space method was used for $\alpha
a^{2}<20$ and the real space method otherwise.

The spatial integrals were evaluated in a limited volume. Specifically, in a
homogeneous system, one has that 
\[
\frac{1}{V}\int d\overrightarrow{r}\;\beta \phi \left( \left\{ n_{\alpha
}\left( \overrightarrow{r}\right) \right\} \right) =\int_{cell}d%
\overrightarrow{r}\;\beta \phi \left( \left\{ n_{\alpha }\left( 
\overrightarrow{r}\right) \right\} \right) /\int_{cell}d\overrightarrow{r}\; 
\]%
where the integrals on the right are restricted to the conventional unit
cell. For the FCC lattice, elementary symmetry considerations show that the
integral on the right can be restricted to the volume $x,y,z>0$ and $x\geq y$
with a symmetry factor of $16$. Further restrictions are possible\cite%
{Groh_Mulder} but were not used. The integrals were then evaluated using an
evenly spaced grid of $20$ points in each direction ( e.g $x=n\delta x/21$
for $0\leq n\leq 20$) . Increasing the number of points had no significant
effect on the calculations.

These calculations are still time consuming, especially at intermediate
values of $\alpha $, so both $\beta F\left( \Gamma \right) $ and $%
K_{ij}^{ab}\left( \Gamma \right) $ were evaluated over a grid of points in
parameter space and bi-cubic spline interpolation\cite{NR} used in the
subsequent calculations. Note that since all of the elementary density
functionals are linear in the average density, it is possible to perform the
calculations for fixed $\alpha $ but many values of the average density at
once, i.e. in parallel, thus saving time. One problem is that these
quantities, especially $K_{ij}^{ab}\left( \Gamma \right) $, are divergent at
high average densities since, for sufficiently high average density, the
local packing fraction $\eta \left( \overrightarrow{r}\right) $can exceed
one which is just the signal of overlapping hard-spheres. For a homogeneous
crystal, the value of the local density for which this divergence occurs is
given by $\overline{\rho }_{latt}/\overline{\eta }\left( 0\right) $, where $%
\overline{\eta(r)}$ is defined in Appendix \ref{app-details} and given
explicitly in eq.(\ref{etabar}). This maximum density is always greater than 
$\overline{\rho}_{latt}$ and approaches $\overline{\rho} _{latt}$ in the
limit $\alpha \rightarrow \infty $. Since most of the variation as a
function of $\overline{\rho }$ occurs near the maximum value, the procedure
used was to discretize the density as%
\[
\overline{\rho }_{k}=\left\{ 
\begin{array}{c}
k\rho _{\max }/k_{\max },\;\;k<k_{\max }/2 \\ 
\rho _{\max }-\frac{1}{2}\rho _{\max }\exp \left( \left( \frac{2k-k_{\max }}{%
k_{\max }}\right) \right) ,\;\;k>=k_{\max }/2%
\end{array}%
\right. 
\]%
with $\rho _{\max }$ chosen to be slightly below $\overline{\rho }_{latt}/%
\overline{\eta }\left( 0\right) $. In the calculations reported below, 200
points were used for the density and the crystallinity was sampled on an
evenly spaced grid of 80 points in the range $0\leq m\leq 0.95$.

Some analytic checks on the numerical calculations are possible. Note that
the density can be written as%
\begin{eqnarray}
\rho \left( \overrightarrow{r}\right) &=&\overline{\rho }\sum_{j}\exp \left(
-k_{j}^{2}/4\alpha \right) \exp \left( i\overrightarrow{K}_{j}\cdot 
\overrightarrow{r}\right) \\
&=&\overline{\rho }\sum_{j}m^{\left( k_{j}^{2}/k_{1}^{2}\right) }\exp \left(
i\overrightarrow{K}_{j}\cdot \overrightarrow{r}\right)  \nonumber \\
&=&\overline{\rho }+\overline{\rho }m\sum_{j\in S_{1}}\exp \left( i%
\overrightarrow{K}_{j}\cdot \overrightarrow{r}\right) +\overline{\rho }%
m^{\left( k_{2}^{2}/k_{1}^{2}\right) }\sum_{j\in S_{s}}\exp \left( i%
\overrightarrow{K}_{j}\cdot \overrightarrow{r}\right) +....  \nonumber
\end{eqnarray}%
where $S_{n}$ is the set of n-th shell reciprocal lattice vectors. It
immediately follows that, in the liquid limit, one has 
\begin{eqnarray}
\lim_{m\rightarrow 0}\frac{\partial \rho \left( \overrightarrow{r};\Gamma
\right) }{\partial \overline{\rho }} &=&1 \\
\lim_{m\rightarrow 0}\frac{\partial \rho \left( \overrightarrow{r};\Gamma
\right) }{\partial m} &=&\overline{\rho }\sum_{j\in S_{1}}\exp \left( i%
\overrightarrow{K}_{j}\cdot \overrightarrow{r}\right) .  \nonumber
\end{eqnarray}%
giving%
\begin{eqnarray}
\lim_{m\rightarrow 0}g_{\rho \rho }\left( \Gamma \right) &=&\frac{2}{3}\pi
\int_{0}^{\infty }r^{4}c_{2}^{l}\left( r;\overline{\rho }\right) dr
\label{g-lim} \\
\lim_{m\rightarrow 0}g_{\rho m}\left( \Gamma \right) &=&0  \nonumber \\
\lim_{m\rightarrow 0}g_{mm}\left( \Gamma \right) &=&\frac{2}{3}\pi \overline{%
\rho }^{2}N_{1}\int_{0}^{\infty }r^{4}c_{2}^{l}\left( r;\overline{\rho }%
\right) \frac{\sin k_{1}r}{k_{1}r}dr.  \nonumber
\end{eqnarray}%
where $c_{2}^{l}\left( r_{12};\overline{\rho }\right) $ is the direct
correlation function of the liquid, $k_{1}$ is the magnitude of the radius
of the first shell of the reciprocal lattice and $N_{1}$ is the number of
elements in the first shell. For an FCC\ lattice in real space, the
reciprocal lattice is BCC giving $N_{1}=8$ and $K_{1}=\left( \frac{4\pi }{a}%
\right) \sqrt{3/4}$. With the RLST theory, the dcf of the liquid is just the
Percus-Yevik dcf, 
\begin{equation}
c_{2}^{PY}\left( r_{12};\overline{\rho }\right) =\left( a_{0}^{py}+a_{1}^{py}%
\frac{r}{\sigma }+a_{2}^{py}\left( \frac{r}{\sigma }\right) ^{3}\right)
\Theta \left( r-\sigma \right) ,
\end{equation}%
with coefficients%
\begin{eqnarray}
a_{0}^{py} &=&-\frac{\left( 1+2\eta \right) ^{2}}{\left( 1-\eta \right) ^{4}}
\\
a_{1}^{py} &=&6\eta \frac{\left( 1+2\eta \right) ^{2}}{\left( 1-\eta \right)
^{4}}  \nonumber \\
a_{2}^{py} &=&\frac{1}{2}\eta a_{0}  \nonumber
\end{eqnarray}%
whereas the White Bear functional gives the same functional form but with
coefficients\cite{WhiteBear}%
\begin{eqnarray}
a_{0}^{wb} &=&-\frac{1+\eta \left( 4+\eta \left( 3-2\eta \right) \right) }{%
\left( 1-\eta \right) ^{4}} \\
a_{1}^{wb} &=&\left( \frac{2-\eta +14\eta ^{2}-6\eta ^{3}}{\left( 1-\eta
\right) ^{4}}+\frac{2\ln \left( 1-\eta \right) }{\eta }\right)  \nonumber \\
a_{2}^{wb} &=&-\left( \frac{3+5\eta \left( \eta -2\right) \left( 1-\eta
\right) }{\left( 1-\eta \right) ^{4}}+\frac{2\ln \left( 1-\eta \right) }{%
\eta }\right) .  \label{last}
\end{eqnarray}%
Equations (\ref{g-lim})-(\ref{last}) give a useful check on the full
numerical calculation.

Except for the cases explicitly discussed above, all numerical integrals,
minimizations and the solution of ordinary differential equations were
performed using routines from the Gnu Scientific Library\cite{GSL}.
One-dimensionsal minimizations were perfomed using either Brent's method or
bisection while numerical integrals were performed using adaptive
integration (the GSL ``QAGS'' routine\cite{GSL}) with relative and absolute
accuracies set to $10^{-4}$. Multidimensional minimizations were performed
using the Simplex algorithm of Nelder and Mead, see e.g. ref. \cite{NR}, as
implemented in GSL, which was terminated whent the simplex size was smaller
than $10^{-4}$.

\bigskip

\section{Results}

\subsection{Bulk coexistence}

As a baseline for the interfacial calculations, it is necessary to know the
predictions of the various theories for the liquid-solid transition. Using
the Gaussian parametrization for the density, eqs.(\ref{d1})-(\ref{d4}), the
expressions given above for the free energy are evaluated and, in accord
with eq.(\ref{thermo}), minimized with respect to $\alpha $ and the lattice
density $\rho _{latt}$while the average density must be adjusted so as to
satisfy the relation between the free energy and the chemical potential
given in eq.(\ref{thermo}). One minimum is always found at $\alpha =0$
corresponding to the uniform liquid having density $\overline{\rho }%
_{l}\left( \mu \right) $. Another occurs at $\alpha _{s}>0$ corresponding to
a bulk solid having some average density $\overline{\rho }_{s}\left( \mu
\right) $. The true equilibrium is whichever of these solutions that
minimizes the grand potential:\ bulk coexistence occurs when they give
identical values%
\begin{equation}
\frac{1}{V}\beta F\left( \overline{\rho }_{l}\left( \mu \right) ,0\right)
-\beta \mu \overline{\rho }_{l}\left( \mu \right) =\frac{1}{V}\beta F\left( 
\overline{\rho }_{s}\left( \mu \right) ,\alpha _{s}\right) -\beta \mu 
\overline{\rho }_{s}\left( \mu \right)
\end{equation}%
or, using eq.(\ref{thermo}) and the definition of the pressure, $\beta P=%
\frac{\partial }{\partial V}\beta F$%
\begin{equation}
\beta P\left( \overline{\rho }_{l}\left( \mu \right),0\right) =\beta P\left( 
\overline{\rho }_{s}\left( \mu \right) ,\alpha _{s}\right)
\end{equation}%
which is the usual condition of coexistence. The results for the RLST and WB
theories are shown in Table \ref{tab1} as well as the values from previous
calculations and the values from simulation. These numbers are on the whole
constant with those in the literature, particularly when it is noted that
the calculations reported in the literature were performed with the
occupancy fixed to be one. The one exception to this general agreement is
the RLST theory where the present results differ noticeably from those
reported by Rosenfeld et al.\cite{Rosenfeld_1997_1}. However, as an
independent check, the evaluations of Warshavsky and Song\cite{Song_FMT} are
also shown and are seen to be consistent with the present calculations thus
supporting the accuracy of the present evaluations.

\begin{table}[tbp]
\caption{The solid density $\overline{\protect\rho }_{sol}\protect\sigma^{3}$%
, liquid density $\overline{\protect\rho }_{liq}\protect\sigma^{3}$, reduced
pressure $\protect\beta P\protect\sigma^{3}$ and chemical potential $\protect%
\mu $ at bulk coexistence as determined from (a) the present work,
(b)Rosenfeld et al \protect\cite{Rosenfeld_1997_1}, (c) Warshavsky and Song %
\protect\cite{Song_FMT}, (d) Roth et al.\protect\cite{WhiteBear} and (e)
from the simulations of Hoover et al\protect\cite{Hoover_1968_1}.}
\label{tab1}%
\begin{ruledtabular}
\begin{tabular}{ccccc}
Theory & $\overline{\rho }_{sol}\sigma^{3}$ & $\overline{\rho }_{liq}\sigma^{3}$ & $\beta P\sigma^{3}$ & $\mu $ \\ \hline
RLST(a) & $1.017$ & $0.935$ & $12.18$ & $16.88$ \\ 
RLST(b) & $1.031$ & $0.938$ & $12.3$ & 17.05 \\ 
RLST(c) & $1.020$ & $0.937$ & $12.26$ & 16.99 \\ 
WB(a) & $1.022$ & $0.932$ & $11.20$ & 15.67 \\ 
WB(c) & $1.023$ & $0.934$ &$11.29$ &  \\ 
WB(d) & $1.023$ & $0.934$ &  &  \\ 
MD(e) & $1.040$ & $0.940$ & $11.70$ & 
\end{tabular}
\end{ruledtabular}
\end{table}

\bigskip

\subsection{The planer interface}

The simplest application of the theory is to the planer liquid-solid
interface. The order parameters are assumed to vary in only one direction,
which is writen as $u=\overrightarrow{r}\cdot \widehat{n}$ where $\widehat{n}
$ is the normal to the interface, from a bulk solid in the limit $%
u\rightarrow -\infty $ to a bulk liquid at $u\rightarrow \infty $. For these
calculations, the GL free energy functional takes the form 
\begin{equation}
\beta \Omega _{GL}^{(Planer)}\left[ \Gamma \right] =A\int_{-\infty }^{\infty
}du\;\left[ \frac{1}{V}\beta F\left( \Gamma \left( u\right) \right) -\beta
\mu \overline{\rho }+\frac{1}{2}\left( g_{ab}\left( \Gamma \left( u\right)
\right) +2h_{ab}\left( \Gamma \left( u\right) \right) \left( \widehat{n}_{x}%
\widehat{n}_{y}+\widehat{n}_{x}\widehat{n}_{z}+\widehat{n}_{y}\widehat{n}%
_{z}\right) \right) \Gamma _{a}^{\prime }\left( u\right) \Gamma _{b}^{\prime
}\left( u\right) \right]  \label{planer}
\end{equation}%
where $A$ is the area in the x-y plane. The calculations were performed for
a fixed lattice structure at the lattice density appropriate for bulk phase
coexistence as determined above.

It is important to realize that the liquid-solid interface is only stable
for a unique value of chemical potential. This is because the interface is
expected to be a localized structure that decays towards the bulk phases
exponentially as one moves away from the interface\cite{GuntonProtein}.
However, since the bulk free energy function, $\beta F\left( \Gamma \right) $%
, and the matrices $g_{ab}\left( \Gamma \right) $ and $h_{ab}\left( \Gamma
\right) $ are being approximated via interpolation of tabulated values, it
is unlikely that the coexistence properties will be exactly the same as
found above. Therefore, a necessary step is to determine coexistence at this
fixed lattice structure using the interpolated $\beta F\left( \Gamma \right) 
$. The procedure used was as follows. Given an initial guess of $\overline{%
\rho }_{sol}$, the value of $\alpha _{sol}$ was determined by finding the
minimum of $\beta F\left( \overline{\rho }_{sol},\alpha \right) $. From
this, the chemical potential is determined from the thermodynamic relation,
eq.(\ref{thermo}) where it is important that the derivative is taken at
constant $\alpha $ and $\overline{\rho} _{latt}$. Then, the liquid density
is determined from the requirement that $\beta F\left( \overline{\rho }%
_{liq},0\right) $ give the same chemical potential. Finally, the bulk
pressures, $\frac{1}{V}\beta F\left( \Gamma \right) -\beta \mu \overline{%
\rho }$, are compared for the liquid and solid and the difference used to
adjust the value of $\overline{\rho }_{sol}$. The process is iterated until
a value of $\overline{\rho }_{sol}$ is found that results in equal pressures
between the bulk phases. The results of the bulk calculations are given in
Table \ref{tab2} where the ``exact'' calculations and those based on the
interpolaton scheme are seen to be in good agreement.

One interesting quantity displayed in this table is the difference between
the lattice density and the average density. The values given for the RLST
theory are in reasonable agreement with the asymptotic values obtained
analytically by Groh\cite{Groh_vacancy}. However, the WB theory gives a
small but negative difference in densities which would mean that instead of
vacancies, the theory predicts interstitials. This surprising and unphysical
result could be an artifact of the WB theory (which, as discussed above, is
an ad hoc extension of Tarazona's theory) or it could be a numerical
artifact indicating that the true vacancy concentration is too small to be
reliably calculated with the numerical methods used. Decreasing the
tolerances of the numerical evaluations in these calculations did not result
in a change of the sign of the vacancy concentration leaving open the
possibility that the result is a real artifact of the theory. Only analytic
work along the lines of ref.\cite{Groh_vacancy} will resolve this question.
Nevertheless, it is important to note that the chemical potential is very
sensitive to the differences between $\overline{\rho }_{sol}$ and $\overline{%
\rho }_{latt}$, or in other words the value of the occupancy. This is
because the chemical potential is given by 
\begin{equation}
\beta \mu = \frac{\partial }{\partial \overline{\rho }_{sol}}\frac{1}{V}%
\beta F\left( \Gamma \right) =\left. \frac{\partial }{\partial \overline{%
\rho }_{sol}}\frac{1}{V}\beta F\left( \Gamma \right) \right| _{\overline{%
\rho }_{latt}}+\frac{\partial \overline{\rho }_{latt}}{\overline{\rho }_{sol}%
}\left. \frac{\partial }{\partial \overline{\rho }_{latt}}\frac{1}{V}\beta
F\left( \Gamma \right) \right| _{\overline{\rho }_{sol}}
\end{equation}%
Working with a fixed value of $\overline{\rho }_{latt}$ means that the
second term on the right is neglected. This is not a problem if the free
energy is stationary with respect to the lattice density as it should be at
thermodynamic equilibrium. However, for typical values of $\alpha $ in the
solid, the maximum average density, defined as that at which the free energy
and all other quantities diverge, is only on the order of $0.0001\%$ above
the lattice density so that the free energy is very sensitive to changes in
the average density and the lattice density. Thus, neglecting the difference
between these and setting $\overline{\rho }_{sol} = \overline{\rho }_{latt}$%
, can lead to errors on the order of $10\%$ in the chemical potential when
evaluated at fixed lattice density even though the free energy itself is
insensitive to this small difference.

\begin{table}[tbp]
\caption{The solid density $\overline{\protect\rho }_{sol}\protect\sigma^{3}$%
, liquid density $\overline{\protect\rho }_{liq}\protect\sigma^{3}$,
difference between lattice and average densities, reduced pressure $\protect%
\beta P\protect\sigma^{3}$ and chemical potential $\protect\mu $ at bulk
coexistence as determined from the ``exact'' calculations and from the
numerical tables with interpolation.}
\label{tab2}%
\begin{ruledtabular}
\begin{tabular}{cccccc}
Theory & $\overline{\rho }_{sol}\sigma^{3}$ & $\overline{\rho }_{liq}\sigma^{3}$ & $\overline{%
\rho }_{latt}\sigma^{3}-\overline{\rho }_{sol}\sigma^{3}$ & $\mu $ & $\beta P\sigma^{3}$ \\ \hline
RLST-exact & $1.017$ & $0.935$ & $4\times 10^{-5}$ & $16.88$ & $%
12.18 $ \\ 
RLST-interpolated & $1.017$ & $0.937$ & $3\times 10^{-5}$ & $16.98$ & $12.27$ \\ 
WB-exact & $1.022$ & $0.932$ & $-7\times 10^{-8}$ & $15.67$ & $11.20$ \\ 
WB-interpolated & $1.021$ & $0.934$ & $-2\times 10^{-8}$ & $15.75$ & $11.28$%
\end{tabular}
\end{ruledtabular}
\end{table}

\begin{figure}[tbp]
\begin{center}
\resizebox{12cm}{!}{
{\includegraphics{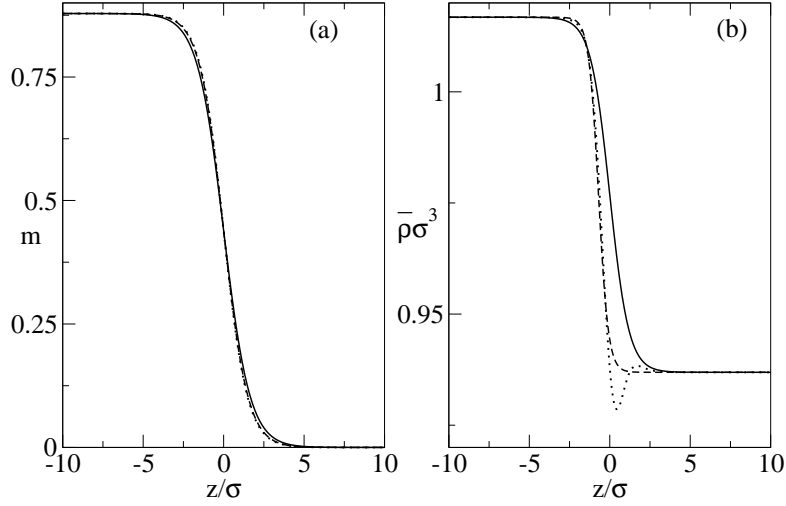}}}
\end{center}
\caption{The crystallinity (a), and the average density (b), profiles as
functions of position perpendicular to the interface as determined using the
RLST DFT and parameterized profiles. Shown are results using hyperbolic
tangent profiles with $B_{\protect\rho} = B_{m}$ (solid lines), allowing $B_{%
\protect\rho} \ne B_{m}$ (dashed lines), and with a Gaussian term in the
density profile (dotted lines). }
\label{fig1}
\end{figure}

\begin{figure}[tbp]
\begin{center}
\resizebox{12cm}{!}{
{\includegraphics{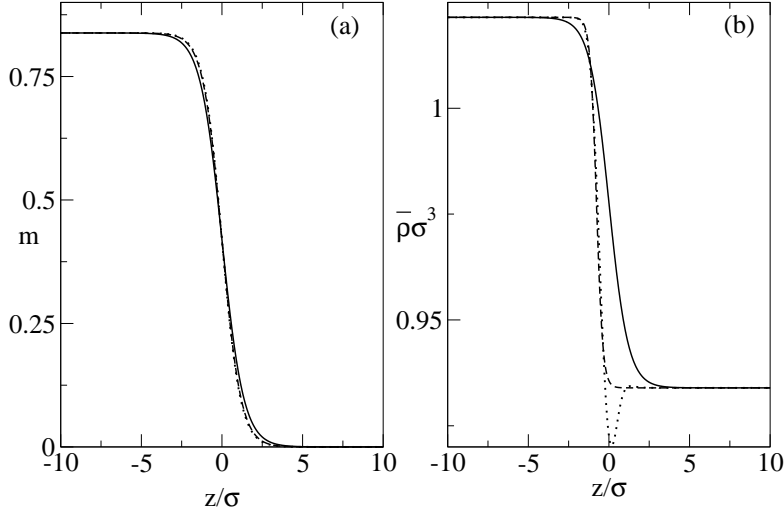}}}
\end{center}
\caption{Same as Fig. \ref{fig1} but using the White Bear DFT.}
\label{fig2}
\end{figure}

\begin{figure}[tbp]
\begin{center}
\resizebox{12cm}{!}{
{\includegraphics{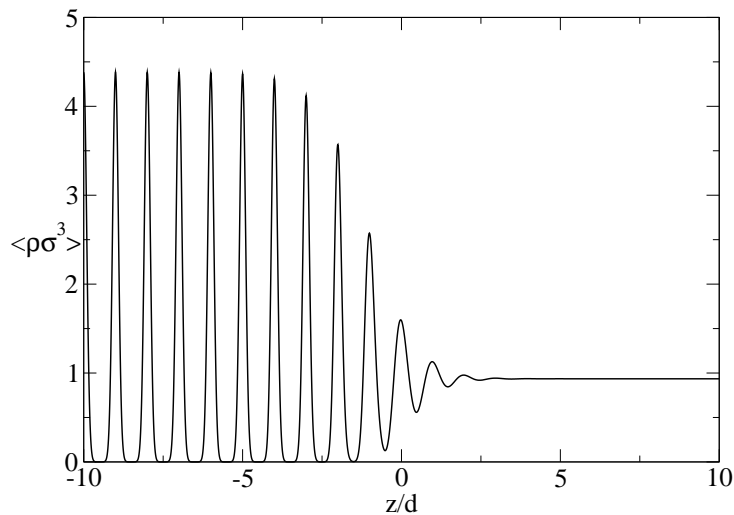}}}
\end{center}
\caption{The atomic density averaged over planes perpendicular to the
interface as a function of position, calculated using the RLST theory and
the offset hyperbolic tangent parameterization. The position is shown in
units of the interplaner spaceing for [100] planes, $d = 0.5a$ where $a$ is
the lattice parameter.}
\label{fig3}
\end{figure}

The structure of the liquid-solid interface is determined by minimizing the
free energy functional, eq.(\ref{planer}), with respect to the
spatially-dependent order parameters leading to the Euler-Lagrange equations%
\begin{equation}
K_{ab}\left( \Gamma \left( u\right) \right) \frac{d^{2}}{du^{2}}\Gamma
_{b}\left( u\right) +\left( \frac{\partial K_{ab}\left( \Gamma \left(
u\right) \right) }{\partial \Gamma _{c}}-\frac{1}{2}\frac{\partial
K_{bc}\left( \Gamma \left( u\right) \right) }{\partial \Gamma _{a}}\right) 
\frac{d\Gamma _{b}\left( u\right) }{du}\frac{d\Gamma _{c}\left( u\right) }{du%
}-\frac{\partial }{\partial \Gamma _{a}}\left( \frac{1}{V}\beta F\left(
\Gamma \left( u\right) \right) -\beta \mu \overline{\rho }\right) =0
\end{equation}%
with%
\[
K_{ab}\left( \Gamma \right) =g_{ab}\left( \Gamma \left( u\right) \right)
+2h_{ab}\left( \Gamma \left( u\right) \right) \left( \widehat{n}_{x}\widehat{%
n}_{y}+\widehat{n}_{x}\widehat{n}_{z}+\widehat{n}_{y}\widehat{n}_{z}\right) 
\]%
and the boundary conditions%
\begin{eqnarray}
\lim_{u\rightarrow -\infty }\Gamma _{a} &=&\Gamma _{a}^{s} \\
\lim_{u\rightarrow -\infty }\Gamma _{a} &=&\Gamma _{a}^{l}  \nonumber
\end{eqnarray}%
where $\Gamma ^{s}=\left( m_{sol},\overline{\rho }_{sol}\right) $, etc.
Direct solution of the Euler-Lagrange equations is difficult due to the
presence of unstable solutions. So, in addition to this method (discussed
below), parameterized forms of the order parameters were explored. Perhaps
the most natural choice is to model the order parameters as simple
hyperbolic tangents,%
\begin{equation}
\Gamma _{a}(u)=\Gamma _{a}^{l}+\left( \Gamma _{a}^{s}-\Gamma _{a}^{l}\right) 
\frac{1}{2}\left(\tanh \left( A_{a}\left( u-B_{a}\right) \right)+1 \right)
\end{equation}%
so that one minimizes the free energy with respect to the widths and
positions of the interface. The surface tension (actually, the surface
free-energy) is calculated as 
\begin{equation}
\tau =\frac{1}{A}\left( \beta \Omega _{GL}^{(Planer)}\left[ \Gamma \right]
-\beta \Omega _{GL}^{bulk}\right)
\end{equation}%
where $\Omega _{GL}^{bulk}$ is the free energy of either bulk phase. The
results are shown in Table \ref{tab3}. First, the simple hyperbolic tangent
profiles with the interface fixed at $B_{m}=B_{\rho }=0$ gives surface
tension of $0.73$ for RLST and $0.75$ for WB. Allowing the interfaces to
move gives a relative displacement of nearly one hard-sphere diameter and
lowers the surface tensions to $\allowbreak 0.67$ and $0.66$ respectively
while giving a considerably narrower density profile. As a further
refinement, inspired by the numerical results discussed below, a Gaussian
term was added giving%
\begin{equation}
\Gamma _{a}=\Gamma _{a}^{l}+\left( \Gamma _{a}^{s}-\Gamma _{a}^{l}\right) 
\frac{1}{2}\left( \tanh \left( A_{a}\left( u-B_{a}\right) \right)+1 \right)
+C_{a}\exp \left( -D_{a}\left( u-E_{a}\right) ^{2}\right) .
\end{equation}%
The addition of the Gaussian makes not appreciable difference for the
profile of the crystallinity, but can affect the profile of the average
density. The profiles of the order parameters are shown for the RLST and WB
theories in Figs.(\ref{fig1}) and (\ref{fig2}) respectively. It is clear
from the figures that there is little change in the profile of the
crystallinity but the density is sensitive to functional forms used. These
results were obtained using as initial values $A_{a}=0.5\sigma $, $%
B_{a}=C_{a}=E_{a}=0$ and $D_{a}=1$. Variation of the initial values by a
factor of two has relatively little effect on the final result. However,.
starting with much sharper widths, $A_{a}>2.5\sigma $, leads in some cases
to a different minimum with a slightly lower surface tension and in other
cases no minimum is found. It could be argued, however, that these cases are
unphysical as the sharpness of the variation of the density violate the
assumptions underlying the derivation of the Ginzburg-Landau form of the
free energy (namely, that the order parameters vary slowly over atomic
distances). The planer density profile, $\rho \left( u\right) =\frac{1}{A}%
\int \rho \left( \overrightarrow{r}\right) \delta \left( u-\overrightarrow{r}%
\cdot \widehat{n}\right) d\overrightarrow{r}$, corresponding to the RLST
result using the combination of hyperbolic tangent and Gaussian is shown in
Fig.(\ref{fig3})) for a profile in the [100] direction. It can be seen that
in agreement with the results reported by\ Warshavsky and Song\cite{Song_FMT}%
, the interface involves perhaps 8 lattice planes. The planer density
corresponding to the offset hyperbolic tangents is indistinguishable from
the one shown. In fact, from Figs. (\ref{fig1}) and (\ref{fig2}) it is clear
that while all three profile functions give indistinguishable results for
the crystallinity, the offset hyperbolic tangents and the hyperbolic tangent
plus Gaussian give very similar profiles for the density and differ from the
simple hyperbolic tangent profile. Furthermore, while in all cases one finds
that going from the bulk liquid to the bulk solid, ordering (i.e. an
increase in crystallinity) precedes densification (an increase in average
density) due to the signficantly greater width of the crystallinity profile,
the more complex parameterizations accentuate this by shifting the
densification curve towards the bulk solid region.

\begin{figure}[tbp]
\begin{center}
\resizebox{12cm}{!}{
{\includegraphics{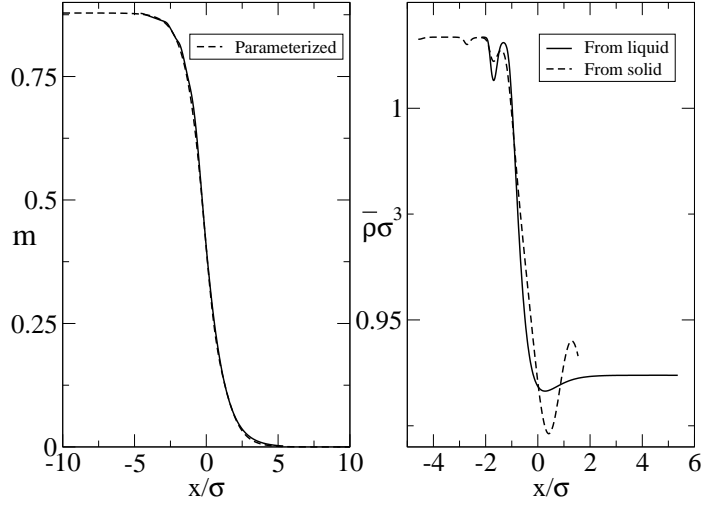}}}
\end{center}
\caption{The crystallinity, left, and the average density, right, obtained
by numerical integration of the Euler-Lagrange equations based on the RLST
DFT. In the case of the crystallinity, the hyperbolic tangent profile
discussed above is also shown. For the average density, the curves obtained
by integrating from both the bulk solid and the bulk liquid regions is
shown. }
\label{fig4}
\end{figure}

The direct integrations of the Euler-Lagrange equations is numerically
challenging as there are both stable,physical solutions and unstable,
unphysical solutions and the accumulation of numerical errors means that
eventually all numerical integrations of the equations become unstable. To
see this, note that far from the interfaces, one expects that the
Euler-Lagrange equations can be linearized about the bulk values of the
order parameters giving%
\begin{equation}
\frac{d^{2}}{du^{2}}\delta \Gamma _{a}-K_{ab}^{-1}\left( \Gamma^{(0)}\right) %
\left[ \frac{\partial ^{2}}{\partial \Gamma _{b}\partial \Gamma _{c}}\frac{1%
}{V}\beta F\left( \Gamma \right) \right] _{\Gamma^{(0)}}\delta \Gamma_{c}=0
\end{equation}%
where $\Gamma _{a}^{(0)}$ are the bulk values of the order parameters and $%
\delta \Gamma _{a}=\Gamma _{a}-\Gamma _{a}^{(0)}$. The possible solutions
are determined by the eigenvalues and eigenvectors of the matrix occurring
in this equation. It turns out that in both the solid and liquid regions,
there is one positive eigenvalue, allowing for decaying solutions, and one
negative eigenvalue, corresponding to undamped oscillations. It is the
mixing of the unphysical oscillatory solution with the decaying solution
that causes numerical difficulties as illustrated in Fig. (\ref{fig4}) for
the RLST\ theory. The figure shows the result of integrating both from the
bulk liquid towards the interface and from the bulk solid towards the
interface with initial conditions chosen to allow for matching the solution
at some point near the interface (thus constructing a shooting-method
solution). Starting from the liquid, it is in fact possible to integrate
almost completely through the interface until, in the solid region,
pollution from the oscillatory solution causes the density to take on
unphysical values. (Of course, unphysical values lie very close to the bulk
value of the average density thus requiring relatively little inaccuracy to
achieve this.) Note that the curve so obtained shows a slight depletion of
the density near the interface as well as some structure on the solid side
of the interface. Both of these effects are very small, given the overall
small change in density between the liquid and solid, and are probably of
little physical consequence. Integrating from the solid side shows similar
effects as well as similar oscillations in the bulk liquid region. The
surface tension of the shooting solution is $\gamma \sigma ^{2}/k_{B}T\simeq
0.67$ which is not much different from that of the curve beginning from the
liquid or solid sides which give $0.64$ and $0.65$ respectively. These
figures are consistent with the surface tension obtained using the
parameterized profiles and support the contention that the parameterizations
are reasonably accurate.

The surface tension of the planer interface has been determined from both
Molecular Dynamics (MD)\cite{davidchak_2000} and Monte Carlo (MC)\
simulations\cite{song_surface_tension}. The MD gives values of $\gamma
\sigma ^{2}/k_{B}T=0.62,0.65$ and $0.58$ for the [100], [110] and [111]
directions respectively while the MC\ gives $\gamma \sigma
^{2}/k_{B}T=0.64,0.62$ and $0.61$. These results show a weak dependence on
the orientation of the interface and the observed variation is not
consistent between the two methods which perhaps indicates that the
asymmetry is very small. The agreement between these values and those
calculated from the GL model are therefore quite good. As expected, the WB
model, which incorporates a better equation of state for the liquid than
does the RLST\ model, seems to be slightly more accurate. The remaining
differences between the calculations and the simulations can at least in
part be attributed to the imposition of an invariant lattice structure.

\begin{table}[tbp]
\caption{The order parameter profile parameters obtained by minimizing the
free energy. The profiles studied are the hyperbolic tangents with $B_{m}=B_{%
\protect\rho}$ (H), the ``offset'' hyperbolic tangents where $B_{m} \ne
B_{rho} $ (OH), and the hyperbolic tangents with a Gaussian term (HG).Also
included are the results from MD simulations of ref \protect\cite%
{davidchak_2000} and the MC simulations of ref.\protect\cite%
{song_surface_tension} . In all cases, the last column gives the surface
tension.}
\label{tab3}%
\begin{ruledtabular}
\begin{tabular}{ccccccccc}
Theory & Profile & $A_{m}$ & $A_{\rho }$ & $B_{\rho }$ & $C_{\rho }$ & $D_{\rho }$ & $%
E_{\rho }$ & $\gamma \sigma ^{2}/k_{B}T$ \\ \hline
RLST & H & 0.61 & 0.83 & * & * & * & * & 0.730 \\ 
RLST & OH & 0.67 & 1.64 & -0.70 & * & * & * & 0.669 \\ 
RLST & HG & 0.68 & 0.99 & * & -0.039 & 1.27 & 0.04 & \textbf{0.667} \\ 
WB & H & 0.74 & 0.84 & * & * & * & * & 0.754 \\ 
WB & OH & 0.85 & 2.54 & -0.78 & * & * & * & 0.659 \\ 
WB & HG & 0.88 & 1.70 & * & -0.06 & 1.97 & -0.21 & \textbf{0.656} \\ 
\textbf{MD}&  &  &  &  &  &  &  & \textbf{0.617} \\ 
\textbf{MC}&  &  &  &  &  &  &  & \textbf{0.623}%
\end{tabular}
\end{ruledtabular}
\end{table}

\subsection{Solid Clusters}

The GL theory can also be used to study the structure of solid clusters
embedded in the liquid. For a spherically symmetric system, the grand
potential becomes%
\begin{equation}
\beta \Omega _{GL}^{(Spherical)}\left[ \Gamma \right] =4\pi \int_{0}^{\infty
}\;\left[ \beta F\left( \Gamma \left( R\right) \right) -\beta \mu \overline{%
\rho }\left( R\right) +\frac{1}{2}g_{ab}\left( \Gamma \left( R\right)
\right) \frac{d\Gamma _{a}\left( R\right) }{dR}\frac{d\Gamma _{b}\left(
R\right) }{dR}\right] R^{2}dR  \label{GL-s}
\end{equation}%
giving the Euler-Lagrange equations%
\begin{equation}
g_{ab}\left( \Gamma \left( R\right) \right) R^{-2}\frac{d}{dR}R^{2}\frac{%
d\Gamma _{b}\left( R\right) }{dR}+\left( \frac{\partial g_{ab}\left( \Gamma
\left( R\right) \right) }{\partial \Gamma _{c}\left( R\right) }-\frac{1}{2}%
\frac{\partial g_{bc}\left( \Gamma \left( R\right) \right) }{\partial \Gamma
_{a}\left( R\right) }\right) \frac{d\Gamma _{b}\left( R\right) }{dR}\frac{%
d\Gamma _{c}\left( R\right) }{dR}-\frac{\partial }{\partial \Gamma
_{a}\left( R\right) }\left( \beta F\left( \Gamma \left( R\right) \right)
-\beta \mu \overline{\rho }\left( R\right) \right) =0.
\end{equation}%
The boundary conditions are 
\begin{eqnarray}
\lim_{R\rightarrow 0}\frac{d\Gamma _{a}\left( R\right) }{dR} &=&0 \\
\lim_{R\rightarrow \infty }\Gamma _{a}\left( R\right)  &=&\Gamma
_{a}^{l}\left( R\right)   \nonumber
\end{eqnarray}%
Note that the first condition ensures that the first derivative along any
line passing through the origin is continuous, e.g. 
\begin{equation}
\lim_{x\uparrow 0}\frac{d\Gamma _{a}\left( R\right) }{dx}=\lim_{x\downarrow
0}\frac{d\Gamma _{a}\left( R\right) }{dx}.
\end{equation}

Classical nucleation theory (CNT) is based on the observation that, to a
first approximation, the excess free energy of a solid cluster of radius $R$
embedded in a large volume of fluid will be 
\begin{equation}
\Delta \Omega ^{cluster}\equiv \Omega ^{cluster}\left[ \Gamma \right]
-\Omega ^{liquid}\sim \frac{4\pi }{3}R^{3}\Delta \omega ^{bulk}+4\pi
R^{2}\gamma   \label{cntx}
\end{equation}%
where $\Delta \omega ^{bulk}$ is the difference in the bulk solid and liquid
free energies per unit volume. The stability of a cluster of a given radius
clearly depends on the value of $\frac{\partial }{\partial R}\Delta \Omega
^{cluster}$. If the liquid is the favored state, so that $\Omega
_{GL}^{solid}>\Omega _{GL}^{liquid}$, then $\frac{\partial }{\partial R}%
\Delta \Omega ^{cluster}>0$ for all $R$ meaning that any cluster will
collapse to $R=0$. If the solid is favored, $\beta \Omega
_{GL}^{solid}<\beta \Omega _{GL}^{liquid}$, clusters smaller than $%
R_{c}\simeq 2\gamma \left( \Delta \omega ^{bulk}\right) ^{-1}$ will collapse
while those with $R>R_{c}$ will be unstable towards unlimited expansion. At
this point, $\frac{\partial ^{2}}{\partial R^{2}}\Delta \Omega
^{cluster}=-8\pi \gamma <0$ so the critical cluster is a maximum in the free
energy with 
\begin{equation}
\Delta \Omega ^{cluster}\left( R_{c}\right) \simeq \frac{16}{3}\pi \frac{%
\gamma ^{3}}{\left( \Delta \omega ^{bulk}\right) ^{2}}.
\end{equation}%
This has been derived under the assumption that the surface tension is
independent of $R$. In fact, as defined above, the surface tension only
applies to conditions of coexistence so that for hard spheres it is a unique
property. Another way of expressing this is that this classical nucleation
model can only apply very close to coexistence. In that case, one can also
write%
\begin{eqnarray}
\Delta \omega ^{bulk} &=&\left( \frac{1}{V}F_{s}\left( \rho _{s}\right) -\mu
\rho _{s}\right) -\left( \frac{1}{V}F_{l}\left( \rho _{l}\right) -\mu \rho
_{l}\right)  \\
&\simeq &\left( \frac{1}{V}F_{s}\left( \rho _{s}^{coex}\right) +\left( \rho
_{s}-\rho _{s}^{coex}\right) \left( \frac{\partial \frac{1}{V}F_{s}\left(
\rho \right) }{\partial \rho }\right) _{\rho _{s}^{coex}}-\mu \rho
_{s}\right)   \nonumber \\
&&-\left( \frac{1}{V}F_{l}\left( \rho _{s}^{coex}\right) +\left( \rho
_{l}-\rho _{s}^{coex}\right) \left( \frac{\partial \frac{1}{V}F_{l}\left(
\rho \right) }{\partial \rho }\right) _{\rho _{l}^{coex}}-\mu \rho
_{s}\right)   \nonumber \\
&=&\left( \frac{1}{V}F_{s}\left( \rho _{s}^{coex}\right) +\left( \rho
_{s}-\rho _{s}^{coex}\right) \mu ^{coex}-\mu \rho _{s}\right) -\left( \frac{1%
}{V}F_{l}\left( \rho _{l}^{coex}\right) +\left( \rho _{l}-\rho
_{l}^{coex}\right) \mu ^{coex}-\mu \rho _{l}\right)   \nonumber \\
&=&\left( -P^{coex}+\left( \mu ^{coex}-\mu \right) \rho _{s}\right) -\left(
-P^{coex}+\left( \mu ^{coex}-\mu \right) \rho _{l}\right)   \nonumber \\
&=&\left( \rho _{s}-\rho _{l}\right) \left( \mu ^{coex}-\mu \right)  
\nonumber
\end{eqnarray}%
giving the well-known expression%
\begin{equation}
\Delta \Omega ^{cluster}\left( R_{c}\right) \simeq \frac{16}{3}\pi \frac{%
\gamma ^{3}}{\left( \rho _{s}-\rho _{l}\right) ^{2}\left( \mu -\mu
^{coex}\right) ^{2}}.
\end{equation}

As in the previous Section, a parametrized form for the order parameters is
considered. It is again assumed that a hyperbolic tangent is a reasonable
guess for the shape of the interface so, taking account of the boundary
conditions suggests using%
\begin{eqnarray}
\Gamma _{a}\left( R\right)  &=&\Gamma _{a}^{l}+\left( \Gamma _{a}\left(
0\right) -\Gamma _{a}^{l}\right) \frac{1+b_{a}R}{1+\left( b_{a}R\right) ^{2}}%
\frac{1-\tanh \left( A_{a}(R-R_{a})\right) }{1-\tanh \left(
-A_{a}R_{a}\right) } \\
&=&\Gamma _{a}^{l}+\left( \Gamma _{a}\left( 0\right) -\Gamma _{a}^{l}\right) 
\frac{1+b_{a}R}{1+\left( b_{a}R\right) ^{2}}\left( \frac{1+\exp \left(
-2A_{a}R_{a}\right) }{1+\exp \left( 2A_{a}\left( R-R_{a}\right) \right) }%
\right)   \nonumber
\end{eqnarray}%
with%
\begin{equation}
b_{a}=\frac{2A_{a}}{e^{2A_{a}R_{a}}+1}.
\end{equation}%
This function has vanishing gradient at $R=0$ and for large $R$ decays as $%
\exp \left( -A_{a}R\right) /R$ which is  the expected asymptotic form\cite%
{GuntonProtein}. Note that the radius of the cluster is determined by $%
R_{\rho }$ and $R_{m}$which are of course not necessarily equal . For a
fixed value of the chemical potential, $\mu $, the values of $\Gamma _{a}^{l}
$ are computed from the known properties of the bulk liquid. Then, the
profile is used in eq.(\ref{GL-s}) and the free energy minimized with
respect to $\Gamma _{a}\left( 0\right) ,A_{m},A_{\rho }$ and $R_{m}$ for a
fixed value of $R_{\rho }$, which is taken to define the cluster size. In
order to avoid unphysical regions, such as $m>1$, auxiliary variables $u$
and $v$ are defined via $\Gamma _{m}(0)=m_{max}\ast \frac{v^{2}}{1+v^{2}}$
and $\Gamma _{\rho }(0)=\rho _{max}-u^{2}$ where $m_{max}$ and $\rho _{max}$
are the maximum values occurring in the tables.

\begin{figure}[tbp]
\begin{center}
\resizebox{12cm}{!}{
{\includegraphics{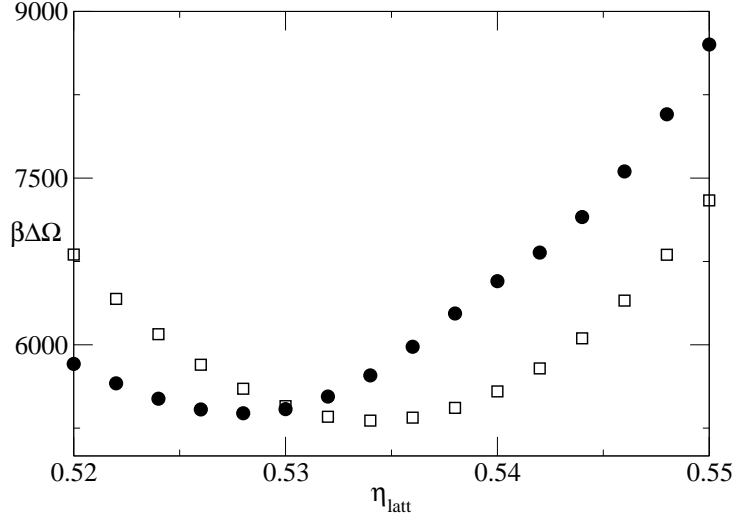}}}
\end{center}
\caption{The excess free energy verses the lattice packing fraction, $%
\protect\eta_{latt}=\protect\pi \protect\rho_{latt} \protect\sigma^{3}/6$
for the RLST theory (filled symbols) and the WB theory (open symbols) for $%
R_{\protect\rho}=30$ and $\protect\beta \Delta \protect\mu = 0.25$}
\label{fig5}
\end{figure}

\begin{figure}[tbp]
\begin{center}
\resizebox{12cm}{!}{
{\includegraphics{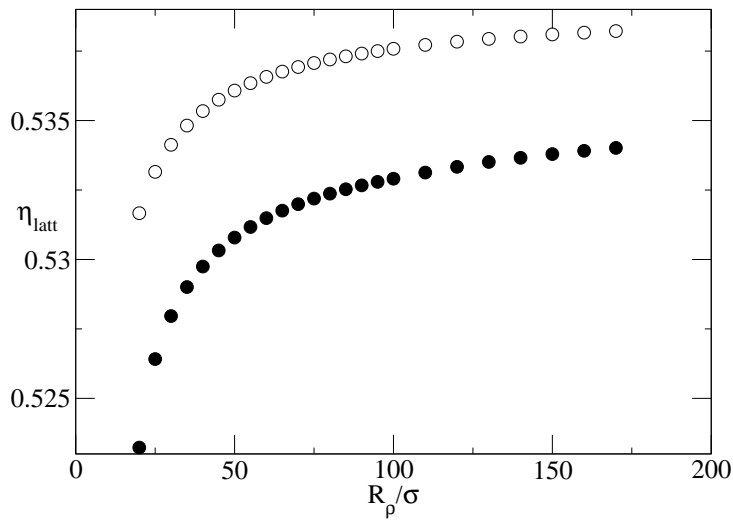}}}
\end{center}
\caption{The lattice packing fraction, $\protect\eta_{latt}=\protect\pi 
\protect\rho_{latt} \protect\sigma^{3}/6$ as a function of the cluster size
for the RLST theory (filled symbols) and the WB theory (open symbols)for $%
\protect\beta \Delta \protect\mu = 0.25$}
\label{fig6}
\end{figure}

The critical cluster then corresponds to the value of $R_{\rho}$ for which
the free energy is stationary. One difference from the planer interface is
that the properties of the solid cluster cannot be assumed to be the same as
those of a bulk solid with the specified chemical potential. Since the
cluster is of finite size, the order parameters at the origin, $\Gamma
_{a}\left( 0\right) $, will in general differ from those of the bulk solid
and, most importantly, the lattice parameter cannot be assumed to be that of
the bulk solid. Thus, in addition to minimizing with respect to the
parameters of the profile, it is also necessary to minimize with respect to
the lattice parameter. These points are illustrated by Fig.\ref{fig5},\
showing the excess free energy as a function of the lattice density for
particular values of the cluster radius and supersaturation. As expected,
the excess free energy shows a minimum for a particular lattice density.
Figure \ref{fig6} shows the variation of the resulting lattice densities as
a function of the radius of the cluster. For small clusters, the density is
significantly lower than that of a bulk solid at the same chemical potential
and increases rapidly as a function of cluster size. For larger clusters,
the rate of change decreases as the bulk limit is approached although the
variation with cluster size is still noticeable even for the largest
clusters ($R_{1}=170\sigma$). The WB theory gives slightly higher densities
than the RLST theory, as would be expected from the bulk coexistence data
(see Table \ref{tab1}). In all cases, the average density at the core, $\bar{%
\rho}(0)$, is very close to the lattice density.

\begin{figure}[tbp]
\begin{center}
\resizebox{12cm}{!}{
{\includegraphics{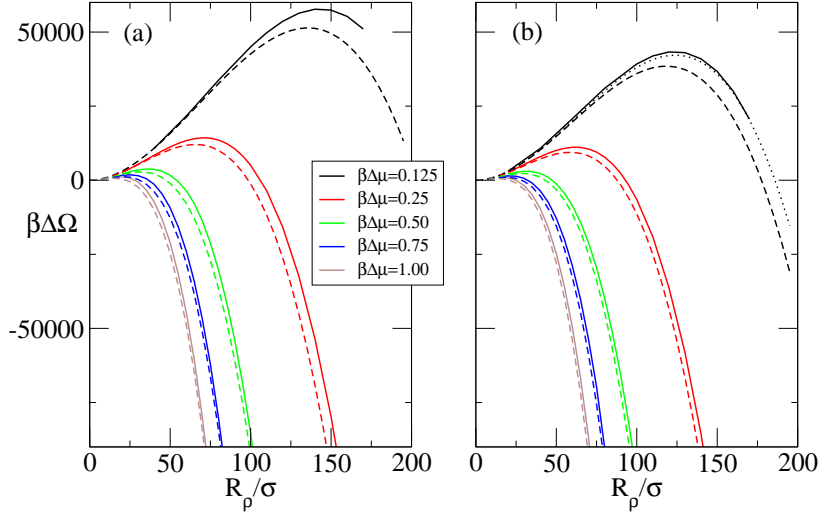}}}
\end{center}
\caption{Excess free energy as a function of the cluster size as determined
using (a) the RLST theory and (b) the WB theory. Curves are shown for $%
\protect\beta \Delta \protect\mu = 0.125$, upper curves, to $\protect\beta %
\Delta \protect\mu = 1.00$, lowest curves. The predicted excess free
energies from CNT are also shown (broken lines).}
\label{fig7}
\end{figure}

Figures \ref{fig7}(a) and \ref{fig7}(b) show the excess free energy as a
function of the cluster size for different values of the supersaturation
determined using the RLST and WB theories respectively. Also shown are the
predicted excess free energy based on CNT, eq. (\ref{cntx}), using the
surface tension at coexistence obtained from the planer calculations of the
previous Section. Clearly, CNT gives a reasonable approximation to the
structure and energy of the critical cluster. In fact, this is a very
sensitive test of the agreement between CNT and the detailed calculations.
The largest discrepancy observed in the figures occurs for the lowest
supersaturation, $\beta \Delta \mu = 0.125$ and at largest cluster sizes.
This is surprising since large clusters approach the bulk limit while
surface tension becomes less important and one would expect that the CNT
calculation would become increasingly accurate. However, since the radius of
the critical cluster diverges as $\beta \Delta \mu \rightarrow 0$, the CNT
results are very sensitive to the value of the chemical potential for small $%
\beta \Delta \mu$. This is illustrated in Fig. \ref{fig7}(b) where nearly
perfect agreement between GL-DFT at $\beta \Delta \mu = 0.125$ is found with
the CNT result for $\beta \Delta \mu = 0.120$ thus suggesting that the
differences seen are at least in part due to numerical inaccuracies in the
determination of the chemical potentials at coexistence. Other factors
contributing to the disagreement are that the CNT assumes that the cluster
has the properties of the bulk solid and that the surface tension is that
for bulk coexisting ($\beta \Delta \mu = 0$) phases.

\begin{figure}[tbp]
\begin{center}
\resizebox{12cm}{!}{
{\includegraphics{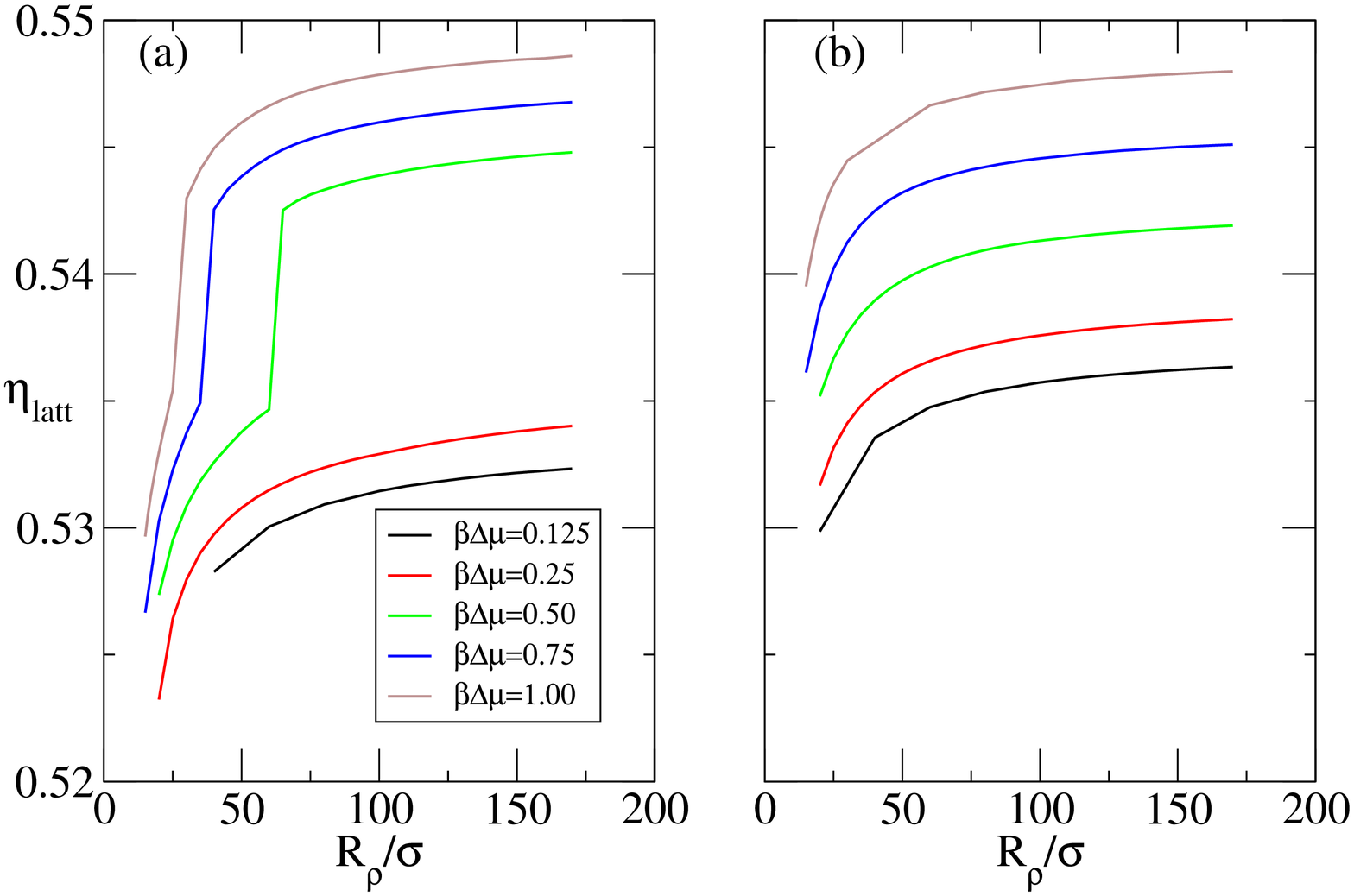}}}
\end{center}
\caption{The lattice packing fraction as a function of the cluster size as
determined using (a) the RLST theory and (b) the WB theory. Curves are shown
for $\protect\beta \Delta \protect\mu = 0.125$, upper curves, to $\protect%
\beta \Delta \protect\mu = 1.00$, lowest curves.}
\label{fig8}
\end{figure}

Figure \ref{fig8} shows that the lattice densities for small clusters are
well below that of the bulk lattice density. A significant difference
between the two DFT's is apparent in that the RLST theory gives a
discontinuity in the lattice density as a function of cluster size while the
WB theory does not. The discontinuity arises because the free energy as a
function of lattice density (for fixed cluster size and supersaturation)
calculated using the RLST theory has two minima. For small clusters, the
low-density minimum dominates while for larger clusters, the high-density
minimum has lowest free energy. Further calculations have verified that for
the lowest supersaturations shown in Fig. \ref{fig8}, the high-density
minimum does dominate for sufficiently large clusters. Furthermore, the
calculations also confirm that the high-density minimum is the one for which
the density and crystallinity are closest to that of a bulk solid.

Finally, in all cases, the structure of the liquid-solid interface is
consistent with the results for the planer interface with similar values for
the widths and relative displacements of the crystallinity and average
density curves.

\section{Conclusions}

The primary results of this paper are the formulation of the Ginzburg-Landau
free energy functional for hard-spheres based on Fundamental Measure Theory
and the use of this functional to study the properties of the liquid-solid
interface and of small FCC solid clusters in solution. It was shown that the
required elements of the GL free energy functional can be calculated
reasonably efficiently using FMT provided that the family of functionals of
the density are extended. The resulting functional was used to study
liquid-solid coexistence with a planer interface as well as the structure of
small solid clusters.

For the planer interface between coexisting liquid and solid phases, the
resulting surface tension is in reasonable agreement with simulation. No
dependence of the surface tension on the lattice plane was found which is
also consistent with simulation. Furthermore, the results using
parameterized profiles and obtained via direct integration of the
Euler-Lagrange equations were found to be consistent, notwithstanding the
numerical difficulties of the latter procedure. This is an important point
as, e.g., the early result obtained by Curtin using the WDA DFT\cite%
{Curtin-Interface}, which preceded determination of the surface tension for
hard-spheres by computer simulation and seemed to be in good agreement with
the later simulations, was subsequently shown by Ohnesorge et al.\cite{OLW2}
to be spurious and due, apparantly, to the use of over-constrained profiles.
That same parameterization was also used by Marr and Gast\cite{Gast1} and,
with some modification, by Kyrlidis and Brown\cite{Kyrlidis_Brown_Surface}.
Recently, Warshavsky and Song, hereafter WS, performed similar calculations
using FMT without the Ginzburg-Landau approximation. They observed greater
differences from simulation than in the present work and non-negligable spatial
asymmetry. The agreement between the present results and simulation may be
fortuitous or the difference between the present results and those of WS may
be due to other details in the later calculation such as the assumption of
unit occupancy. In any case, the agreement between both FMT\ calculations
and simulation is an improvement over the results based on older DFT's which
typically give a value of the planer surface tension that is about half of
that measured in simulation\cite{OLW2,Gast1,Kyrlidis_Brown_Surface}.

The GL functional was also used to study the properties of small solid
clusters in superdense solution. For the range of supersaturtions considered
here, it was not found to be possible to stabilize clusters of radius less
than about 15 hard-sphere radii; the free energy difference from the liquid
is found to be very small and no local minimum in the free energy could be
found. The results for clusters that could be stabilized are consistent with
the predictions of classical nucleation theory. One interesting observation
was that the RLST theory predicts a discontinuity in the structure of small
clusters: very small clusters have unexpectedly low lattice densities while
larger clusters approach the properties of the bulk solid. The crossover
point between the two structures increases as the supersaturation increases
and, conversely, appears to diverge near coexistence. However, since the
RLST is based on a Percus-Yevik description of the fluid, which is not
accurate at such high densities, and since the WB theory does not show this
effect, it seems likely to be an artifact.

For both the planer interface and the clusters, it was found that moving
from the bulk liquid towards the solid, one first observes ordering of the
fluid and then densification. This is interesting as it is the opposite of
the predictions of recent studies of crystallization directly from a low
density gas\cite{lutsko_2006_1}. For the latter case, it seems to always be
more favorable to densify first, thus forming dense liquid droplets, and
then to order. Of course, the main reason that this scenario is not observed
here is that for hard-spheres it is only possible to study nucleation of the
solid from an already-dense fluid - there is no equivalent of the
low-density gas-solid transition. This is also undoubtedly one of the
reasons that CNT\ seems to work so well for hard-spheres (i.e., because
there is no critical point).

Although the Ginzburg-Landau model is reasonably successful in the
applications described here, there are problems which cannot be discounted.
Most important is that the free energy is unstable with respect to very
sharp interfaces. Such rapid variations in the order parameters are outside
the scope of the GL model, which is based on a gradient expansion, and so
have been avoided here by always starting the minimizations from relatively
slowly varying profiles. Within reasonable bounds for the starting
parameters, the resulting profiles are then relatively insensitive to the
starting point.

While of some intrinsic interest in testing the GL model and the density
functional theory, these results are only intended as baselines for more
interesting studies of realistic interaction models such as the
Lennard-Jones potential. Since the only input to the GL model is a
reasonable model for the bulk free energy and a reasonable model for the
bulk direct correlation function, rather than a full blown density
functional theory, it is expected that the extension of this work to other
potentials will be relatively straightforward.

\bigskip

\begin{acknowledgements}
I am grateful to Xeyeu Song for several stimulating discussions on this topic.
This work was supported in part by the European Space Agency/PRODEX
under contract number C90105.
\end{acknowledgements}

\bigskip

\appendix

\section{Details of the calculations}

\label{app-details}

\subsection{The density functionals}

\bigskip All of the linear density functionals are of the form%
\begin{equation}
n\left( \overrightarrow{r};\left[ \rho \right] \right) =\int d%
\overrightarrow{r}_{1}\;w\left( \overrightarrow{r}-\overrightarrow{r}%
_{1}\right) \rho \left( \overrightarrow{r}_{1}\right) .
\end{equation}%
As long as the density is written as a sum of basis functions at each
lattice site,%
\begin{equation}
\rho \left( \overrightarrow{r}\right) =\sum_{n}\overline{\rho }\left( 
\overrightarrow{r}-\overrightarrow{R}_{n}\right)
\end{equation}%
then the functionals can also be written as%
\begin{equation}
n\left( \overrightarrow{r};\left[ \rho \right] \right) =\sum_{n}\overline{n}%
\left( \overrightarrow{r}-\overrightarrow{R}_{n};\left[ \rho \right] \right)
\end{equation}%
with%
\begin{equation}
\overline{n}\left( \overrightarrow{r};\left[ \rho \right] \right) =\int d%
\overrightarrow{r}_{1}\;w\left( \overrightarrow{r}-\overrightarrow{r}%
_{1}\right) \overline{\rho }\left( \overrightarrow{r}_{1}\right) .
\end{equation}%
Before proceeding, note that in the case of tensorial quantities, $\overline{%
T}_{ij...k}\left( \overrightarrow{r};\left[ \rho \right] \right) $, the only
vector available is $\widehat{r}$ and the only tensor, aside from $\widehat{r%
}\widehat{r}$, is the unit tensor. Thus, it follows that 
\begin{eqnarray}
\overline{v}_{i}\left( \overrightarrow{r}\right) &=&v\left( r\right) 
\widehat{r}_{i} \\
\overline{T}_{ij}\left( \overrightarrow{r}\right) &=&A\left( r\right) \delta
_{ij}+B\left( r\right) \widehat{r}_{i}\widehat{r}_{j}  \nonumber \\
\overline{T}_{ijl}\left( \overrightarrow{r}\right) &=&C(r)\left( \widehat{r}%
_{i}\delta _{jl}+\widehat{r}_{j}\delta _{il}+\widehat{r}_{l}\delta
_{ij}\right) +D(r)\widehat{r}_{i}\widehat{r}_{j}\widehat{r}_{l}  \nonumber \\
\overline{T}_{ijlm}\left( \overrightarrow{r}\right) &=&E\left( r\right)
\left( \delta _{ij}\delta _{lm}+\delta _{il}\delta _{jm}+\delta _{im}\delta
_{jl}\right)  \nonumber \\
&&+F\left( r\right) \left( \widehat{r}_{i}\widehat{r}_{j}\delta _{lm}+%
\widehat{r}_{i}\widehat{r}_{l}\delta _{jm}+\widehat{r}_{i}\widehat{r}%
_{m}\delta _{jl}+\widehat{r}_{j}\widehat{r}_{l}\delta _{im}+\widehat{r}_{j}%
\widehat{r}_{m}\delta _{il}+\widehat{r}_{l}\widehat{r}_{m}\delta _{ij}\right)
\nonumber \\
&&+G\left( r\right) \widehat{r}_{i}\widehat{r}_{j}\widehat{r}_{l}\widehat{r}%
_{m}  \nonumber
\end{eqnarray}%
The basic functionals for FMT are then found to be%
\begin{eqnarray}
\overline{s}\left( \overrightarrow{r}\right) &=&\frac{1}{2r}\sqrt{\frac{%
\alpha \sigma ^{2}}{\pi }}\left( \exp \left( -\alpha \left( r-\frac{\sigma }{%
2}\right) ^{2}\right) -\exp \left( -\alpha \left( r+\frac{\sigma }{2}\right)
^{2}\right) \right)  \label{etabar} \\
\overline{\eta }\left( \overrightarrow{r}\right) &=&\frac{1}{2}\left(
erf\left( \sqrt{\alpha }\left( r+\frac{\sigma }{2}\right) \right)
\allowbreak -erf\left( \sqrt{\alpha }\left( r-\frac{\sigma }{2}\right)
\right) \right) \allowbreak \allowbreak -\frac{1}{\alpha \sigma }\overline{s}%
\left( \overrightarrow{r}\right)  \nonumber \\
\overline{v}\left( \overrightarrow{r}\right) &=&\frac{1}{2r}\allowbreak 
\sqrt{\frac{\alpha \sigma ^{2}}{\pi }}\left( \exp \left( -\alpha \left( r-%
\frac{\sigma }{2}\right) ^{2}\right) +\exp \left( -\alpha \left( r+\frac{%
\sigma }{2}\right) ^{2}\right) \right) -\frac{1}{\sigma \alpha r}\overline{s}%
\left( \overrightarrow{r}\right)  \nonumber
\end{eqnarray}%
and the quantities needed for the tensorial functionals are%
\begin{eqnarray}
A\left( r\right) &=&\frac{1}{\sigma \alpha r}v\left( r\right) \\
B(r) &=&\overline{s}\left( r\right) -\frac{3}{\sigma \alpha r}v\left(
r\right) \allowbreak  \nonumber \\
C\left( r\right) &=&\frac{1}{\alpha r\sigma }\allowbreak \allowbreak B\left(
r\right)  \nonumber \\
D\left( r\right) &=&\overline{v}\left( r\right) -5C\allowbreak \left(
r\right)  \nonumber \\
E\left( r\right) &=&\frac{1}{\alpha ^{2}\sigma ^{2}r^{2}}B\left( r\right) 
\nonumber \\
F\left( r\right) &=&\frac{1}{\alpha r\sigma }\overline{v}\left( r\right)
-5E\left( r\right)  \nonumber \\
G\left( r\right) &=&B\left( r\right) -7F\left( r\right)  \nonumber
\end{eqnarray}%
Finally, one needs the additional quantities%
\begin{eqnarray}
\overline{\eta }^{i}\left( \overrightarrow{r}\right) &=&H\left( r\right) 
\widehat{r}_{i} \\
\overline{\eta }^{ij}\left( \overrightarrow{r}\right) &=&I\left( r\right) 
\widehat{r}_{i}\widehat{r}_{j}+J\left( r\right) \delta _{ij}  \nonumber
\end{eqnarray}%
with%
\begin{eqnarray}
H\left( r\right) &=&\allowbreak r\overline{\eta }\left( r\right) -\frac{1}{%
2\alpha }\overline{v}\left( r\right) \allowbreak \allowbreak \\
I\left( r\right) &=&rH\left( r\right) -\frac{\sigma }{4\alpha }B(r) 
\nonumber \\
J\left( r\right) &=&\frac{1}{2\alpha r}H\left( r\right) .  \nonumber
\end{eqnarray}

In Fourier space one has that%
\begin{eqnarray}
n_{\alpha }\left( \overrightarrow{r};\left[ \rho \right] \right)
&=&\sum_{j}\rho _{j}\exp \left( i\overrightarrow{K}_{j}\cdot \overrightarrow{%
r}_{1}\right) \overline{n}_{\alpha }\left( \overrightarrow{K}_{j}\right) \\
\overline{n}_{\alpha }\left( \overrightarrow{K}\right) &=&\int d%
\overrightarrow{r}_{2}\;\exp \left( -i\overrightarrow{K}\cdot 
\overrightarrow{r}_{2}\right) w_{\alpha }\left( \overrightarrow{r}%
_{2}\right) .  \nonumber
\end{eqnarray}%
The tensorial quantities can be expressed as in real space but with the
vector $\widehat{r}$ replaced by $\widehat{k}$ so%
\begin{eqnarray}
\overline{v}_{i}\left( \overrightarrow{K}\right) &=&v\left( r\right) 
\widehat{k}_{i} \\
\overline{T}_{ij}\left( \overrightarrow{K}\right) &=&A\left( k\right) \delta
_{ij}+B\left( k\right) \widehat{k}_{i}\widehat{k}_{j}  \nonumber \\
\overline{T}_{ijl}\left( \overrightarrow{K}\right) &=&C(k)\left( \widehat{k}%
_{i}\delta _{jl}+\widehat{k}_{j}\delta _{il}+\widehat{k}_{l}\delta
_{ij}\right) +D(k)\widehat{k}_{i}\widehat{k}_{j}\widehat{k}_{l}  \nonumber \\
\overline{T}_{ijlm}\left( \overrightarrow{K}\right) &=&E\left( k\right)
\left( \delta _{ij}\delta _{lm}+\delta _{il}\delta _{jm}+\delta _{im}\delta
_{jl}\right)  \nonumber \\
&&+F\left( k\right) \left( \widehat{k}_{i}\widehat{k}_{j}\delta _{lm}+%
\widehat{k}_{i}\widehat{k}_{l}\delta _{jm}+\widehat{k}_{i}\widehat{k}%
_{m}\delta _{jl}+\widehat{k}_{j}\widehat{k}_{l}\delta _{im}+\widehat{k}_{j}%
\widehat{k}_{m}\delta _{il}+\widehat{k}_{l}\widehat{k}_{m}\delta _{ij}\right)
\nonumber \\
&&+G\left( k\right) \widehat{k}_{i}\widehat{k}_{j}\widehat{k}_{l}\widehat{k}%
_{m}  \nonumber
\end{eqnarray}%
The scalar and vector functionals are 
\begin{eqnarray}
\overline{s}\left( k\right) &=&\pi \sigma ^{2}j_{0}\left( \frac{k\sigma }{2}%
\right) \\
\overline{\eta }\left( k\right) &=&\frac{1}{6}\pi \sigma ^{3}\left(
j_{0}\left( \frac{1}{2}k\sigma \right) +j_{2}\left( \frac{1}{2}k\sigma
\right) \right)  \nonumber \\
\overline{v}_{a}\left( k\right) &=&-i\pi \sigma ^{2}\frac{k_{a}}{k}%
j_{1}\left( \frac{k\sigma }{2}\right)  \nonumber
\end{eqnarray}%
and the coefficients for the tensorial functionals are%
\begin{eqnarray}
A\left( k\right) &=&\frac{\pi \sigma ^{2}}{3}\left( j_{0}\left( \frac{%
k\sigma }{2}\right) +j_{2}\left( \frac{k\sigma }{2}\right) \right) \\
B\left( k\right) &=&-\pi \sigma ^{2}j_{2}\left( \frac{k\sigma }{2}\right) 
\nonumber \\
C(k) &=&-i\frac{1}{5}\sigma ^{2}\pi \left( j_{1}\left( \frac{k\sigma }{2}%
\right) +j_{3}\left( \frac{k\sigma }{2}\right) \right)  \nonumber \\
D(k) &=&i\sigma ^{2}\pi j_{3}\left( \frac{k\sigma }{2}\right)  \nonumber \\
E\left( k\right) &=&\frac{1}{105}\sigma ^{2}\pi \left( 7j_{0}\left( \frac{%
k\sigma }{2}\right) +10j_{2}\left( \frac{k\sigma }{2}\right) +3j_{4}\left( 
\frac{k\sigma }{2}\right) \right)  \nonumber \\
F\left( k\right) &=&-\frac{1}{7}\sigma ^{2}\pi \left( j_{2}\left( \frac{%
k\sigma }{2}\right) +j_{4}\left( \frac{k\sigma }{2}\right) \right)  \nonumber
\\
G\left( k\right) &=&\sigma ^{2}\pi j_{4}\left( \frac{k\sigma }{2}\right) 
\nonumber
\end{eqnarray}%
Finally, 
\begin{eqnarray}
\overline{\eta }^{i}\left( \overrightarrow{K}\right) &=&H\left( k\right) 
\widehat{k}_{i} \\
\overline{\eta }^{ij}\left( \overrightarrow{K}\right) &=&I\left( k\right) 
\widehat{k}_{i}\widehat{k}_{j}+J\left( k\right) \delta _{ij}  \nonumber
\end{eqnarray}%
with%
\begin{eqnarray}
H\left( k\right) &=&i\frac{\sigma ^{4}\pi }{20}\left( j_{1}\left( \frac{1}{2}%
k\sigma \right) +j_{3}\left( \frac{1}{2}k\sigma \right) \right) \allowbreak
\allowbreak \\
I\left( k\right) &=&-\frac{\pi \sigma ^{5}}{56}\left( j_{2}\left( \frac{1}{2}%
k\sigma \right) +j_{4}\left( \frac{1}{2}k\sigma \right) \right)  \nonumber \\
J\left( k\right) &=&\frac{\pi \sigma ^{5}}{840}\left( 7j_{0}\left( \frac{1}{2%
}k\sigma \right) +10j_{2}\left( \frac{1}{2}k\sigma \right) +3j_{4}\left( 
\frac{1}{2}k\sigma \right) \right) .  \nonumber
\end{eqnarray}

\subsection{The derivatives of the free energy}

The free energy is written in terms of the integral of 
\[
\phi \left( \left\{ n_{\alpha }\left( \overrightarrow{r}\right) \right\}
\right) =\phi _{1}\left( \left\{ n_{\alpha }\left( \overrightarrow{r}\right)
\right\} \right) +\phi _{2}\left( \left\{ n_{\alpha }\left( \overrightarrow{r%
}\right) \right\} \right) +\phi _{3}\left( \left\{ n_{\alpha }\left( 
\overrightarrow{r}\right) \right\} \right) 
\]%
and for the gradient term in the GL functional one needs the second
derivatives of this function. Consider the first two terms which are the
same for all of the DFTs,%
\begin{equation}
\phi _{1}+\phi _{2}=-\frac{1}{\pi \sigma ^{2}}s\ln \left( 1-\eta \right) +%
\frac{1}{2\pi \sigma }\frac{s^{2}-v^{2}}{\left( 1-\eta \right) }.
\end{equation}%
The second derivatives of these are%
\begin{eqnarray}
\frac{\partial ^{2}}{\partial \eta ^{2}}\left( \phi _{1}+\phi _{2}\right) &=&%
\frac{1}{\pi \sigma ^{2}}\frac{s}{\left( 1-\eta \right) ^{2}}+\frac{1}{\pi
\sigma }\frac{s^{2}-v^{2}}{\left( 1-\eta \right) ^{3}}  \label{rlst1} \\
\frac{\partial ^{2}}{\partial s^{2}}\left( \phi _{1}+\phi _{2}\right) &=&%
\frac{1}{\pi \sigma }\frac{1}{\left( 1-\eta \right) }  \nonumber \\
\frac{\partial ^{2}}{\partial v_{i}\partial v_{j}}\left( \phi _{1}+\phi
_{2}\right) &=&\frac{1}{\pi \sigma }\frac{-1}{\left( 1-\eta \right) }\delta
_{ij}  \nonumber
\end{eqnarray}%
and the cross-derivatives are%
\begin{eqnarray}
\frac{\partial ^{2}}{\partial \eta \partial s}\left( \phi _{1}+\phi
_{2}\right) &=&\frac{1}{\pi \sigma ^{2}}\frac{1}{1-\eta }+\frac{1}{\pi
\sigma }\frac{s}{\left( 1-\eta \right) ^{2}}  \label{rlst2} \\
\frac{\partial ^{2}}{\partial v_{i}\partial \eta }\left( \phi _{1}+\phi
_{2}\right) &=&\frac{1}{\pi \sigma }\frac{-v_{i}}{\left( 1-\eta \right) ^{2}}
\nonumber \\
\frac{\partial ^{2}}{\partial v_{i}\partial s}\left( \phi _{1}+\phi
_{2}\right) &=&0.  \nonumber
\end{eqnarray}%
The third term takes different forms depending on the theory. For the RLST\
theory%
\begin{equation}
\phi _{3}^{RLST}=\frac{1}{24\pi }\frac{s^{3}}{\left( 1-\eta \right) ^{2}}%
\left( 1-\frac{v^{2}}{s^{2}}\right) ^{3}
\end{equation}%
and the second derivatives are%
\begin{eqnarray}
\frac{\partial ^{2}}{\partial \eta ^{2}}\phi _{3}^{RLST} &=&\frac{1}{4\pi }%
\frac{s^{3}}{\left( 1-\eta \right) ^{4}}\left( 1-\frac{v^{2}}{s^{2}}\right)
^{3}  \label{rlst3} \\
\frac{\partial ^{2}}{\partial s^{2}}\phi _{3}^{RLST} &=&\frac{1}{4\pi }\frac{%
s}{\left( 1-\eta \right) ^{2}}\left( 1-\frac{v^{2}}{s^{2}}\right) \left( 1+%
\frac{v^{2}}{s^{2}}+2\left( \frac{v^{2}}{s^{2}}\right) ^{2}\right)  \nonumber
\\
\frac{\partial ^{2}}{\partial v_{i}\partial v_{j}}\phi _{3}^{RLST} &=&\frac{1%
}{4\pi }\frac{s^{3}}{\left( 1-\eta \right) ^{2}}\left( 1-\frac{v^{2}}{s^{2}}%
\right) \left( \frac{4v_{i}v_{j}+\delta _{ij}\left( v^{2}-s^{2}\right) }{%
s^{4}}\right)  \nonumber
\end{eqnarray}%
and the cross derivatives are%
\begin{eqnarray}
\frac{\partial ^{2}}{\partial \eta \partial s}\phi _{3}^{RLST} &=&\frac{1}{%
4\pi }\frac{s^{2}}{\left( 1-\eta \right) ^{3}}\left( 1-\frac{v^{2}}{s^{2}}%
\right) ^{2}\left( 1+\frac{v^{2}}{s^{2}}\right)  \label{rlst4} \\
\frac{\partial ^{2}}{\partial \eta \partial v_{i}}\phi _{3}^{RLST} &=&-\frac{%
1}{2\pi }\frac{sv_{i}}{\left( 1-\eta \right) ^{3}}\left( 1-\frac{v^{2}}{s^{2}%
}\right) ^{2}  \nonumber \\
\frac{\partial ^{2}}{\partial s\partial v_{i}}\phi _{3}^{RLST} &=&-\frac{1}{%
4\pi }\frac{v_{i}}{\left( 1-\eta \right) ^{2}}\left( 1-\frac{v^{2}}{s^{2}}%
\right) \left( 1+3\frac{v^{2}}{s^{2}}\right)  \nonumber
\end{eqnarray}

For the WB theory, first consider the simpler Tarazona theory which has 
\begin{equation}
\phi _{3}^{T}=\frac{3}{16\pi }\frac{\overrightarrow{v}\cdot 
\overleftrightarrow{T}\cdot \overrightarrow{v}-sv^{2}-Tr\left( 
\overleftrightarrow{T}^{3}\right) +sTr\left( \overleftrightarrow{T}%
^{2}\right) }{\left( 1-\eta \right) ^{2}}
\end{equation}%
so%
\begin{eqnarray}
\frac{\partial ^{2}}{\partial \eta ^{2}}\phi _{3}^{T} &=&\frac{6}{\left(
1-\eta \right) ^{2}}\phi _{3}^{T} \\
\frac{\partial ^{2}}{\partial s^{2}}\phi _{3}^{T} &=&0  \nonumber \\
\frac{\partial ^{2}}{\partial v_{i}\partial v_{j}}\phi _{3}^{T} &=&\frac{3}{%
16\pi }\frac{T_{ij}+T_{ji}-2s\delta _{ij}}{\left( 1-\eta \right) ^{2}}=\frac{%
3}{8\pi }\frac{T_{ij}-s\delta _{ij}}{\left( 1-\eta \right) ^{2}}  \nonumber
\\
\frac{\partial ^{2}}{\partial T_{ij}\partial T_{lm}}\phi _{3}^{T} &=&\frac{3%
}{16\pi }\frac{-3\delta _{jl}T_{mi}-3T_{jl}\delta _{il}+2s\delta _{lj}\delta
_{mi}}{\left( 1-\eta \right) ^{2}}  \nonumber
\end{eqnarray}%
and the cross-derivatives are%
\begin{eqnarray}
\frac{\partial ^{2}}{\partial \eta \partial s}\phi _{3}^{T} &=&\frac{3}{8\pi 
}\frac{-v^{2}+Tr\left( \overleftrightarrow{T}^{2}\right) }{\left( 1-\eta
\right) ^{3}} \\
\frac{\partial ^{2}}{\partial \eta \partial v_{i}}\phi _{3}^{T} &=&\frac{3}{%
4\pi }\frac{T_{ia}v_{a}-sv_{i}}{\left( 1-\eta \right) ^{3}}=\frac{3}{4\pi }%
\frac{\left( T_{ia}-s\delta _{ia}\right) v_{a}}{\left( 1-\eta \right) ^{3}} 
\nonumber \\
\frac{\partial ^{2}}{\partial \eta \partial T_{ij}}\phi _{3}^{T} &=&\frac{%
v_{i}v_{j}-3T_{jc}T_{ci}+2sT_{ji}}{\left( 1-\eta \right) ^{3}}  \nonumber \\
\frac{\partial ^{2}}{\partial s\partial v_{i}}\phi _{3}^{T} &=&-\frac{3}{%
8\pi }\frac{v_{i}}{\left( 1-\eta \right) ^{2}}  \nonumber \\
\frac{\partial ^{2}}{\partial s\partial T_{ij}}\phi _{3}^{T} &=&\frac{3}{%
8\pi }\frac{T_{ji}}{\left( 1-\eta \right) ^{2}}  \nonumber \\
\frac{\partial ^{2}}{\partial T_{ij}\partial v_{l}}\phi _{3}^{T} &=&\frac{3}{%
16\pi }\frac{\delta _{il}v_{j}+\delta _{jl}v_{i}}{\left( 1-\eta \right) ^{2}}
\nonumber
\end{eqnarray}%
Now, in the WB theory one has 
\begin{eqnarray}
\phi _{3}^{WB} &=&F\left( \eta \right) \phi _{3}^{T} \\
F\left( \eta \right) &=&\frac{2}{3}\frac{\eta +\left( 1-\eta \right) ^{2}\ln
\left( 1-\eta \right) }{\eta ^{2}}  \nonumber \\
F^{\prime }\left( \eta \right) &=&\frac{2}{3}\frac{\eta ^{2}-2\eta -2\left(
1-\eta \right) \ln \left( 1-\eta \right) }{\eta ^{3}}  \nonumber \\
F^{\prime \prime }\left( \eta \right) &=&\frac{2}{3}\frac{6\eta -\eta
^{2}+\left( 6-4\eta \right) \ln \left( 1-\eta \right) }{\eta ^{4}}  \nonumber
\end{eqnarray}%
Then, 
\begin{eqnarray}
\frac{\partial ^{2}}{\partial \Gamma _{a}\partial \Gamma _{b}}\phi _{3}^{WB}
&=&F\left( \eta \right) \frac{\partial ^{2}}{\partial \Gamma _{a}\partial
\Gamma _{b}}\phi ^{T} \\
&&+\delta _{b\eta }F^{\prime }\left( \eta \right) \frac{\partial }{\partial
\Gamma _{a}}\phi ^{T}+\delta _{a\eta }F^{\prime }\left( \eta \right) \frac{%
\partial }{\partial \Gamma _{b}}\phi ^{T}  \nonumber \\
&&+\delta _{a\eta }\delta _{b\eta }F^{\prime \prime }\left( \eta \right)
\phi ^{T}  \nonumber
\end{eqnarray}%
or%
\begin{equation}
\frac{\partial ^{2}}{\partial \Gamma _{a}\partial \Gamma _{b}}\phi
_{3}^{WB}=\left( F\left( \eta \right) +\frac{1}{2}\left( 1-\eta \right)
\left( \delta _{a\eta }+\delta _{b\eta }\right) F^{\prime }\left( \eta
\right) \right) \frac{\partial ^{2}}{\partial \Gamma _{a}\partial \Gamma _{b}%
}\phi ^{T}+\delta _{a\eta }\delta _{b\eta }F^{\prime \prime }\left( \eta
\right) \phi ^{T}
\end{equation}

\section{Vanishing of $h_{ab}$ for the RLST\ theory}

\label{App-h}

Combining eqs.(\ref{gh}) and (\ref{c2}) gives

\begin{equation}
h_{ab}\left( \Gamma \right) =-\frac{1}{4V}\sum_{\alpha ,\beta }\int d%
\overrightarrow{r}_{1}d\overrightarrow{r}_{2}d\overrightarrow{r}%
\;\;r_{12x}r_{12y}\frac{\partial ^{2}\phi \left( \left\{ n_{\alpha }\left( 
\overrightarrow{r}\right) \right\} \right) }{\partial n_{\alpha }\partial
n_{\beta }}w_{\alpha }\left( \overrightarrow{r}-\overrightarrow{r}%
_{1}\right) w_{\beta }\left( \overrightarrow{r}-\overrightarrow{r}%
_{2}\right) \frac{\partial \rho \left( \overrightarrow{r}_{1};\Gamma \right) 
}{\partial \Gamma _{a}}\frac{\partial \rho \left( \overrightarrow{r}%
_{2};\Gamma \right) }{\partial \Gamma _{b}}.  \label{app1}
\end{equation}%
The goal here is to prove that this quantity vanishes in the RLST theory.
The idea behind the proof is to make a change of variables in the integral
whereby $\overrightarrow{r}_{i}=\left( x_{i},y_{i},z_{i}\right) \rightarrow 
\overrightarrow{r}_{i}^{\prime }=\left( -x_{i},y_{i},z_{i}\right) $ for $%
i=1,2$. This clearly gives an overall sign change as well as affecting the
arguments of the various functions occurring under the integral. Using the
fundamental fact that for a uniform solid (i.e. spatially constant $\Gamma $%
) with a cubic lattice structure, the density has reflection symmetry so
that $\rho \left( \overrightarrow{r}_{i};\Gamma \right) =\rho \left( 
\overrightarrow{r}_{i}^{\prime };\Gamma \right) $ , it follows that the low
order density functionals occurring in the RLST theory have simple parity
and it is therefore possible to show that, aside from the overall change of
sign, the integrand is invariant. This proves that $h_{ab}\left( \Gamma
\right) =0$.

To begin, 
\begin{eqnarray}
h_{ab}\left( \Gamma \right) &=&-\frac{1}{4V}\sum_{\alpha ,\beta }\int d%
\overrightarrow{r}_{1}d\overrightarrow{r}_{2}d\overrightarrow{r}%
\;\;x_{12}y_{12}\frac{\partial ^{2}\phi \left( \left\{ n_{\alpha }\left( 
\overrightarrow{r}\right) \right\} \right) }{\partial n_{\alpha }\partial
n_{\beta }}w_{\alpha }\left( \overrightarrow{r}-\overrightarrow{r}%
_{1}\right) w_{\beta }\left( \overrightarrow{r}-\overrightarrow{r}%
_{2}\right) \frac{\partial \rho \left( \overrightarrow{r}_{1};\Gamma \right) 
}{\partial \Gamma _{a}}\frac{\partial \rho \left( \overrightarrow{r}%
_{2};\Gamma \right) }{\partial \Gamma _{b}}  \label{app2} \\
&=&\frac{1}{4V}\sum_{\alpha ,\beta }\int d\overrightarrow{r}_{1}^{\prime }d%
\overrightarrow{r}_{2}^{\prime }d\overrightarrow{r}\;\;x_{12}y_{12}\frac{%
\partial ^{2}\phi \left( \left\{ n_{\alpha }\left( \overrightarrow{r}\right)
\right\} \right) }{\partial n_{\alpha }\partial n_{\beta }}w_{\alpha }\left( 
\overrightarrow{r}-\overrightarrow{r}_{1}^{\prime }\right) w_{\beta }\left( 
\overrightarrow{r}-\overrightarrow{r}_{2}^{\prime }\right) \frac{\partial
\rho \left( \overrightarrow{r}_{1}^{\prime };\Gamma \right) }{\partial
\Gamma _{a}}\frac{\partial \rho \left( \overrightarrow{r}_{2}^{\prime
};\Gamma \right) }{\partial \Gamma _{b}}  \nonumber \\
&=&\frac{1}{4V}\sum_{\alpha ,\beta }\int d\overrightarrow{r}_{1}^{\prime }d%
\overrightarrow{r}_{2}^{\prime }d\overrightarrow{r}^{\prime }\;\;x_{12}y_{12}%
\frac{\partial ^{2}\phi \left( \left\{ n_{\alpha }\left( \overrightarrow{r}%
^{\prime }\right) \right\} \right) }{\partial n_{\alpha }\partial n_{\beta }}%
w_{\alpha }\left( \overrightarrow{r}^{\prime }-\overrightarrow{r}%
_{1}^{\prime }\right) w_{\beta }\left( \overrightarrow{r}^{\prime }-%
\overrightarrow{r}_{2}^{\prime }\right) \frac{\partial \rho \left( 
\overrightarrow{r}_{1}^{\prime };\Gamma \right) }{\partial \Gamma _{a}}\frac{%
\partial \rho \left( \overrightarrow{r}_{2}^{\prime };\Gamma \right) }{%
\partial \Gamma _{b}}.  \nonumber
\end{eqnarray}%
where we have also made the change of variable $\overrightarrow{r}=\left(
x,y,z\right) \rightarrow \overrightarrow{r}^{\prime }=\left( -x,y,z\right) $%
. The claim is that $\Delta =0,$ where 
\begin{eqnarray*}
\Delta &=&\sum_{\alpha ,\beta }\Delta _{\alpha \beta } \\
\Delta _{\alpha \beta } &=&\frac{\partial ^{2}\phi ^{RLST}\left( \left\{
n_{\alpha }\left( \overrightarrow{r}^{\prime }\right) \right\} \right) }{%
\partial n_{\alpha }\partial n_{\beta }}w_{\alpha }\left( \overrightarrow{r}%
^{\prime }-\overrightarrow{r}_{1}^{\prime }\right) w_{\beta }\left( 
\overrightarrow{r}^{\prime }-\overrightarrow{r}_{2}^{\prime }\right) \\
&&-\frac{\partial ^{2}\phi ^{RLST}\left( \left\{ n_{\alpha }\left( 
\overrightarrow{r}\right) \right\} \right) }{\partial n_{\alpha }\partial
n_{\beta }}w_{\alpha }\left( \overrightarrow{r}-\overrightarrow{r}%
_{1}\right) w_{\beta }\left( \overrightarrow{r}-\overrightarrow{r}_{2}\right)
\end{eqnarray*}%
Now, the scalar weight functions depend only on the magnitude of the
separation, e.g. $w_{\eta }\left( \delta \overrightarrow{r}^{\prime }\right)
=w_{\eta }\left( \left| \delta \overrightarrow{r}^{\prime }\right| \right)
=w_{\eta }\left( \left| \delta \overrightarrow{r}\right| \right) $, so that
they are invariant under the change of variables. The vector density $w_{%
\overrightarrow{v}}\left( \left| \delta \overrightarrow{r}^{\prime }\right|
\right) =$ $\left( \delta \overrightarrow{r}^{\prime }/\left| \delta 
\overrightarrow{r}^{\prime }\right| \right) \Theta \left( \left| \delta 
\overrightarrow{r}^{\prime }\right| -\frac{\sigma }{2}\right) =$ $\left(
\delta \overrightarrow{r}^{\prime }/\left| \delta \overrightarrow{r}\right|
\right) \Theta \left( \left| \delta \overrightarrow{r}\right| -\frac{\sigma 
}{2}\right) $ so that $w_{v_{x}}\left( \left| \delta \overrightarrow{r}%
^{\prime }\right| \right) $ has odd parity and the other components have
even parity. From this and the reflection symmetry of $\rho \left( 
\overrightarrow{r};\Gamma \right) $, it follows that the density functionals 
$\eta \left( \overrightarrow{r}\right) $ , $s\left( \overrightarrow{r}%
\right) $ and $\overrightarrow{v}\left( \overrightarrow{r}\right) $ have the
corresponding parities (that is , all are even except $v_{x}\left( 
\overrightarrow{r}^{\prime }\right) =-v_{x}\left( \overrightarrow{r}\right) $%
). From the explicit expressions for $\frac{\partial ^{2}\phi ^{RLST}}{%
\partial n_{\alpha }\partial n_{\beta }}$ given in the previous appendix, it
is seen that all of these have even parity except for those in which one and
only one of the derivatives is with respect to $v_{x}\left( \overrightarrow{r%
}\right) $. It immediately follows that 
\[
\Delta _{ss}=\Delta _{\eta \eta }=\Delta _{s\eta }=\Delta _{\eta s}=\Delta
_{v_{x}v_{x}}=\Delta _{v_{y}v_{y}}=\Delta _{v_{z}v_{z}}=\Delta
_{v_{y}v_{z}}=0. 
\]%
Finally, $\frac{\partial ^{2}\phi ^{RLST}}{\partial s\partial v_{x}},$ $%
\frac{\partial ^{2}\phi ^{RLST}}{\partial \eta \partial v_{x}},\frac{%
\partial ^{2}\phi ^{RLST}}{\partial v_{x}\partial v_{y}}$ and $\frac{%
\partial ^{2}\phi ^{RLST}}{\partial v_{x}\partial v_{z}}$ are all of odd
parity (because they are proportional to $v_{x}$) as are $%
w_{s}w_{v_{x}},w_{\eta }w_{v_{x}},w_{v_{x}}w_{v_{y}}$ and $%
w_{v_{x}}w_{v_{z}} $ so that $\Delta _{sv_{x}}=\Delta _{\eta v_{x}}=\Delta
_{v_{z}v_{y}}=\Delta _{v_{z}v_{z}}=0$ thus proving that $\Delta =0$ and
hence that $h_{ab}\left( \Gamma \right) =0$ in the RLST theory. The same is
not true of the WB theory as terms such as $T_{jc}T_{ci}$, and therefore 
\[
\frac{\partial ^{2}}{\partial \eta \partial T_{ij}}\phi _{3}^{T}=\frac{%
v_{i}v_{j}-3T_{jc}T_{ci}+2sT_{ji}}{\left( 1-\eta \right) ^{3}}, 
\]%
do not have a definite parity under reflection.

\bigskip 
\bibliographystyle{apsrev}
\bibliography{../physics}

\end{document}